\documentclass[12pt,oneside]{article}
\usepackage[dvips]{graphicx}
\usepackage{amsmath,epsfig,amssymb,graphicx}
\usepackage{ntheorem}
\usepackage{fullpage}
\usepackage[footnotesize,bf]{caption}
\theoremstyle{plain}
\theoremheaderfont{\sc}
\theorembodyfont{\upshape}
\theoremstyle{nonumberplain}
\theoremseparator{}
\theoremsymbol{\rule{1ex}{2ex}}

\newcommand{\Pm}{\mathbf{P}}
\newcommand{\wm}{\mathbf{W}}
\newcommand{\mm}{\mathbf{M}}
\newcommand{\tm}{\mathbf{T}}

\newcommand{\Pc}{\mathcal{P}}

\begin{document}
\title{On Cyclic and Nearly Cyclic Multiagent Interactions\\ in the Plane}
\author{Fr\'ed\'erique Oggier and Alfred
Bruckstein\thanks{Visiting Professor from The Technion - IIT, Haifa,
Israel.}\\
\\
Division of Mathematical Sciences \\
School of Physical and Mathematical Sciences \\
Nanyang Technological University, Singapore.\\
{\tt frederique@ntu.edu.sg, freddy@cs.technion.ac.il}. 
}

\maketitle \abstract{We discuss certain types of cyclic and nearly cyclic
interactions among $N$ ``point"-agents in the plane, leading
to formations of interesting limiting geometric configurations.
Cyclic pursuit and local averaging interactions have been analyzed in 
the context of multi-agent gathering. In this paper, we consider some 
nearly cyclic interactions that break symmetry leading to factor circulants 
rather than circulant interaction matrices.
}

%
%

\section{Introduction} 

Consider a ``swarm" or ``pack" of $N$ robots in
the plane, denoted by $ \mathcal{P}_0, \mathcal{P}_1, \ldots
\mathcal{P}_{N-1}$ which can all see each other and are aware of the
other robot's identities (i.e., can distinguish them). We shall
define the rules of interaction specifying how each robot
$\mathcal{P}_k$ moves in response to the (evolution in time of the)
configuration of the entire swarm. Therefore denoting
$\mathcal{P}_k$'s location at time $t$ to be
$\mathcal{P}_k(t)=x_k(t)+iy_k(t)$ (a complex number), we assume that
we can write the swarm evolution equations as follows:
\begin{eqnarray}
\frac{d\mathcal{P}_k(t)}{dt}&=&
\Phi_k^{(C)}\{\mathcal{P}_s(\xi)|_{s=0,1,\ldots N-1}; \xi\leq t\} \nonumber\\
\textrm{or}  \ \  \mathcal{P}_k(t+1)&=&
\Phi_k^{(D)}\{\mathcal{P}_s(\xi)|_{s=0,1,\ldots N-1}; \xi\leq t\}\label{eq:eq1}
\end{eqnarray}
depending on whether the temporal evolution is continuous $(C)$ or
discrete $(D)$. So far the $\Phi$-operators
are not specified, and in fact they could be quite involved in
general. 
The operator $\Phi_k^{(C)}$ provides an instantaneous velocity vector for 
agent $\Pc_k$ in response to the locations of the other agents in the swarm, 
while $\Phi_k^{(D)}$ will yield the next location for $\Pc_k$ in a 
synchronous discrete timed evolution. These operators should produce the 
same motion if we decide to look at the agents in different frames 
of reference, i.e., re-encode their locations using transformed coordinates, 
hence the resulting equations should be at least similarity invariant, and 
maybe even affine invariant. The requirement to have the same evolution 
equation in arbitrarily similarity (i.e., scaled Euclidean) or affine 
transformed coordinates clearly imposes restrictions on the $\Phi$ operators 
and some of these will be discussed in the sequel.  

An important class of operators are the linear memoryless
ones which have the form
\begin{equation}
\Phi_k \{\mathcal{P}_0, \mathcal{P}_1,\ldots
\mathcal{P}_{N-1}\}=\sum_{l=0}^{N-1} m^k_l(t)\mathcal{P}_l(t) \nonumber
\end{equation}
where $m_l^k(t)$ are some (complex) numbers, varying perhaps in
time. In this case, Equation (\ref{eq:eq1}) describes a linear
(generally time varying) system's state evolution, and there is a
wealth of theory dealing with such systems in the control and signal
processing literature. Here we shall mainly be concerned with a
special class of (constant) linear Toeplitz operators of the form
\begin{eqnarray}
\Phi_k \{\mathcal{P}_0, \mathcal{P}_1,\ldots
\mathcal{P}_{N-1}\}=\sum_{l=0}^{N-1}
\lambda^{\textrm{Ind}[(l-k)<0]}m_{(l-k)\textrm{mod}~N}
\mathcal{P}_l(t)
\end{eqnarray}\label{eq:eq2}
where $\lambda$ is some complex number, and
\begin{displaymath}
\left\{
\begin{array}{c}
m_{-1}\equiv m_{N-1}\textrm{mod}~N \\
 m_{-k}\equiv m_{N-k}\textrm{mod}~N
\end{array}\right.
\textrm{and} \ \ 
 \textrm{Ind} [a<0]=\left\{\begin{array}{lll}1& if & a<0\\
0 & if & a \geq 0
\end{array}\right. .
\end{displaymath}

Writing out explicitly $\Phi_k\{\Pc_0,\ldots,\Pc_{N-1}\}$ for 
$k=0,\ldots,N-1$ in matrix form and denoting 
\[
\Pm(t)=
\left[
\begin{array}{c}
\Pc_0(t)\\
\vdots\\
\Pc_{N-1}(t)
\end{array}
\right],
\] 
the swarm's evolution dynamics becomes
\begin{eqnarray}
\left(\frac{d}{dt}\Pm(t) ~\textrm{or}
\right)\Pm(t+1) 
&=&
\Phi \Pm(t) \\
&=&\left[\begin{array}{ccccl}
m_0 & m_1& m_2 & \ldots& m_{N-1}\\
\lambda m_{N-1}& m_0& m_1& \ldots& m_{N-2}\\
\lambda m_{N-2}& \lambda m_{N-1}& m_0&\ldots& \ldots\\
\vdots & \vdots & \ldots &\ddots& \ldots\\
\lambda m_1 & \lambda m_2 & \ldots & \lambda m_{N-1}& m_0
\end{array}
\right]\Pm(t).\nonumber
\end{eqnarray}
Note here that if $\lambda = 1$, the matrix is a special
Toeplitz-circulant matrix, otherwise it is a generalization 
of a circulant called a $\lambda$-factor, or $\lambda$-circulant 
matrix. Such matrices arise in several applications, such as linear 
systems theory \cite{GO,HS}, linear algebra \cite{B}, geometry 
\cite{D,S1,S2}, and in connection with inverses of Toeplitz matrices 
\cite{F,G,CPW}, coding theory \cite{EOK} and linear systems 
of differential equations \cite{W}. In case of $\lambda = 1$,
i.e., when the operator $\Phi$ is Toeplitz-circulant, we have that 
all the robotic agents perform ``cyclically" the same
operation, i.e. agent $\mathcal{P}_k$ will determine its next
location (or its velocity) according to the same weighted average
performed on $ \mathcal{P}_k, \mathcal{P}_{k+1}, \ldots
\mathcal{P}_{(k+N)\textrm{mod}N}$ (in this order), i.e.

\begin{eqnarray}
\left\{
\begin{array}{l}
\mathcal{P}_k(t+1)\\
\textrm{or } \frac{d}{dt}\mathcal{P}_k(t)
 \end{array}
\right\}
&=&
[m_0, m_1, \ldots m_{N-1}]\left[
\begin{array}{l}
\mathcal{P}_k(t)\\
\mathcal{P}_{k+1}(t)\\
\vdots\\
 \mathcal{P}_{(k+N)\textrm{mod}~N}(t)
 \end{array}
 \right]
\end{eqnarray}
which can be rewritten as
\begin{eqnarray}
\left\{
\begin{array}{l}
\mathcal{P}_k(t+1)\\
\textrm{or } \frac{d}{dt}\mathcal{P}_k(t)
 \end{array}
\right\}
 & = &
\bar{m}\left[
\begin{array}{ccrccccc}
0&0&0&1&^{(k^{th}place)}&\ldots&0\\
&&&&1&&\\
&&&&&1&\\
&&&&&&\ddots\\
1&&&&&&\\
&1&&&&&\\
&&1&&&\ldots&0\\
0&\ldots&&1&\ldots&0&0
\end{array}
  \right]
\left[
 \begin{array}{l}
\mathcal{P}_0(t)\\
\mathcal{P}_1(t)\\
\vdots\\
 \mathcal{P}_{N-1}(t)
 \end{array}
 \right] \nonumber \\
 &=& \bar{m} \mathbf{Z}^{k-1}\cdot\Pm(t) 
\end{eqnarray}
where  
\[
\mathbf{Z}\triangleq  \left[
\begin{array}{ccccc}
0&1&&&0\\
\vdots&0&1&&\\
\vdots&&0&1&\\
0&&&0&1\\
1&0&\cdots&&0
\end{array}
  \right] \mbox{ and }
\bar{m}=[m_0, m_1, \ldots m_{N-1}].
\]

This special case, with a circulant matrix $\Phi$, was extensively
analyzed before in the context of polygon smoothing evolutions and 
cyclic pursuits for robotic gathering and
formation control, see e.g. \cite{D, S1, S2, BSS, BCE, MBF, MB, F}.

Note that invariance requirements impose some conditions on the linear 
evolution operators, as we now discuss.
If $\Pm(t)$ is described by the evolution equations
\begin{eqnarray*}
\frac{d}{dt}\Pm(t)&=& \Phi^{(C)} \Pm(t)\\
\mbox{or } \Pm(t+1) &=&   \Phi^{(D)}\Pm(t)
\end{eqnarray*}
from some initial location $\Pm(0)=\Pm(t=0)$, and if we re-encode 
the agents' positions via a general similarity transformation of 
the form
\[
\Pm'(t)\triangleq \rho\Pm(t)+\tau {\bf 1}
\]
where $\rho$ and $\tau$ are some complex numbers and 
${\bf 1}=[1,\ldots,1]^T$, we shall have for $\Pm'(t)$:
\begin{itemize}
\item
in the continuous case
\begin{eqnarray*}
\frac{d}{dt}\Pm'(t)
&\triangleq& \frac{d}{dt}\left(\rho\Pm(t)+\tau {\bf 1}\right)\\
&=& \rho \frac{d}{dt}\Pm(t)\\
&=& \rho \Phi^{(C)}\Pm(t)
\end{eqnarray*}
which is equal to $ \Phi^{(C)}\left(\rho\Pm(t)+\tau {\bf 1}\right)$ only if 
$\Phi^{(C)}{\bf 1}={\bf 0} $.
\item
in the discrete case
\begin{eqnarray*}
\Pm'(t+1)
&\triangleq& \rho\Pm(t+1)+\tau {\bf 1}\\
&=& \rho \Phi^{(D)}\Pm(t)+\tau {\bf 1}\\
\end{eqnarray*}
which is equal to $ \Phi^{(D)}\left(\rho\Pm(t)+\tau {\bf 1}\right)$ only if 
$\Phi^{(D)}{\bf 1}={\bf 1} $.
\end{itemize}

Hence the $\Phi$-matrices that describe linear, time-invariant 
evolutions need to obey the conditions $\Phi^{(C)}{\bf 1}={\bf 0}$ or 
$\Phi^{(D)}{\bf 1}={\bf 1}$ in order to have Euclidean or similarity 
invariant evolutions. In some of our examples, these conditions cannot 
be satisfied. However, note that any $N\times N$ matrix $\Phi$ may be 
embedded in an $(N+1)\times (N+1)$ matrix ${\bf \Phi}$ as follows
\[
\left[
\begin{array}{cc}
\Phi & {\bf s} \\
{\bf 0} & z
\end{array}
\right]
\left[
\begin{array}{c}
1 \\
\vdots\\
1
\end{array}
\right]=
\left[
\begin{array}{c}
\Phi{\bf 1}+ {\bf s} \\
z
\end{array}
\right]
\]
and selecting either $z=0$ and ${\bf s}=-\Phi{\bf 1}$ or $z=1$, 
we obtain a ${\bf \Phi}$ matrix that describes an invariant evolution 
of a multi-agent system with an additional agent $\Pc_B$ whose position 
is stationary ($\frac{d}{dt} \Pc_B=0$ or $\Pc_B(t+1)=\Pc_B(t)$). 
This additional agent will act as a ``beacon'' or a set reference point, 
for the description of the swarm of agents. In this case, setting 
$\Pc_B=(0,0)$, the evolution of the rest of the agents will be 
described by the original matrix $\Phi$. 
Note that the spatial location of the fixed $\Pc_B$ in the plane 
may be determined according to the initial location of the agents of 
the swarm. A good example is the geometric and affine invariant decision 
that can be made by each agent independently to set $\Pc_B$, and hence 
the origin of its Cartesian coordinate system, at the centroid of the 
agent location constellation at $t=0$.
This will make the swarm evolution entirely autonomous. However, an 
external setting of the location of $\Pc_B$ might be useful in controlling 
the swarm and steering it toward a desired place in the environment. One 
might even desire to move $\Pc_B$ in time and make the swarm move accordingly, 
by tracking the beacon point in addition to its own internal dynamics
controlled by $\Phi$.

%
%

\section{Analyzing Swarm Evolution via Mode Decoupling}

Circulant, and $\lambda$-factor circulant matrices have very special
structures and this allows us to diagonalize them, essentially by
Fourier transform methods. Let us see, in general, how
diagonalization yields a way to analyze the evolution of the
constellation of robots by decoupling it into independently evolving
modes. Indeed assume that the time-invariant matrix $\Phi$ can be
diagonalized (for example when $\Phi$ has distinct eigenvalues,
hence a full set of orthonormal eigenvectors), as follows
\[
\Phi= T^{-1}DT
\]
where $D=\mbox{Diag}[d_0, d_1\ldots d_{N-1}]$ displays the eigenvalues of
$\Phi$ and the columns of $T^{-1}$ are the (right) eigenvectors. Now
we have that
\[
\left.
\begin{array}{cc}
& \Pm(t+1)\\
\textrm{or} & \\
& \frac{d}{dt}\Pm(t)
\end{array}
\right\} = T^{-1}DT\Pm(t)
\]
and hence
\[
\left.
\begin{array}{cc}
& T\Pm(t+1)\\
&\frac{d}{dt}(T\Pm(t))
\end{array}
\right\} = D(T\Pm(t)).
\]
In terms of the transformed vector $\widetilde{\Pm}(t)\triangleq T \Pm(t)$, 
the evolution is a decoupled evolution controlled explicitly by the 
(constant) eigenvalues \cite{HS}. Indeed, we have
\[
\widetilde{\Pm}(t)=
{\left[
\begin{array}{cccc}
d_o^t&&&\\
& d_1^t&0&\\
&& \ddots & \\
& 0 &&d_{N-1}^t
\end{array}
\right] \widetilde{\Pm}(0)}_{\textrm{(discrete case)}}
\]
or \[
\widetilde{\Pm}(t)=
{\left[
\begin{array}{cccc}
e^{d_0t}&&&\\
& e^{d_1t}&0&\\
&& \ddots & \\
& 0 &&e^{d_Nt}
\end{array}
\right] \widetilde{\Pm}(0)}_{\textrm{(continous case)}}.
\]
Therefore diagonalization enables the explicit solution of the 
swarm evolution, in the case the $\Phi$ matrix is time invariant 
and has a full set of orthonormal eigenvectors. As we shall see below, 
$\lambda$-factor circulants are a family of matrices that enable both a nice 
physical interpretation in terms of cyclic and symmetric interactions 
among similar agents and an explicit diagonalization via 
discrete Fourier transform matrices.

%
%

\section{Diagonalization of Factor Circulants}

Factor circulant matrices are very special in that they provide 
explicit formulae for the diagonalizing transforms and for their 
eigenvalues. This enables us to analyze in detail the behavior of 
multiagent interactions when these are cyclic  or ``nearly" cyclic, 
and fully describe the limiting behaviors of the swarm. 
For circulants, we have the following results.
Consider the unitary Fourier transform matrix
\begin{eqnarray*}
[\mathbf{FT}]
&\triangleq & 
\frac{1}{\sqrt{N}}\left[
\begin{array}{llll}
w^0& w^0& \ldots&w^0\\
w^0 & w^1& \ldots &w^{N-1}\\
\vdots& \vdots&&\vdots\\
w^0 & w^{N-1}& \ldots & w^{(N-1)(N-1)}
\end{array}
\right] \\
&=&
\frac{1}{\sqrt{N}}\left[
w^{(k-1)(l-1)}
\right]_{k,l=1,\ldots,N}
\end{eqnarray*}
where $w=e^{-i\frac{2\pi}{N}}$ is an $N^{th}$ root of unity. 
Then $\mathbf{C}$ is a Toeplitz-circulant matrix if and only if
\[
 \mathbf{C}[\mathbf{FT}]=[\mathbf{FT}]
 \left[
\begin{array}{cccc}
\mu_o&&&\\
& \mu_1&0&\\
&& \ddots & \\
& 0 &&\mu_{N-1}
\end{array}
\right]
\]
where $ \mu_0, \mu_1, \ldots, \mu_{N-1}$ are the eigenvalues 
of $\mathbf{C}$ and are given by
\begin{eqnarray*}
\mu_l=\sum_{k=0}^{N-1}c_k e^{-i\frac{2\pi}{N}kl}. 
\end{eqnarray*}
Hence
\[
[\mathbf{FT}]^* \mathbf{C}[\mathbf{FT}]=
\textrm{Diag}[\mu_0, \mu_1, \ldots, \mu_{N-1}]
\]
and 
\[
\mathbf{C}= [\mathbf{FT}]\textrm{Diag}[\mu_0, \mu_1, \ldots, \mu_{N-1}]
[\mathbf{FT}]^*.
\]

To summarize the remarkable properties of circulants, we can state
that they are (1) diagonalized by the discrete Fourier Transform,
(2) they all commute, (3) their products are circulants, (4) their
sums are circulants too, and (5) their inverses/pseudoinverses are
circulants, and are readily found \cite{G}. In fact, many of the
wonders of modern signal processing algorithms, and linear, time
invariant systems theory stem from the above properties. 

The corresponding, and equally remarkable properties of
$\lambda$-circulants are, however, much less known and applied.
Suppose we consider the following operation on a circulant
$\mathbf{C}=\mathbf{C}_{[c_0, c_1, \ldots, c_{N-1}]}$:
\[
\mathbf{W}= \left[
\begin{array}{cccc}
a_o&&&\\
& a_1&0&\\
&& \ddots & \\
& 0 &&a_{N-1}
\end{array}
\right]
\mathbf{C}_{[c_0, c_1, \ldots, c_{N-1}]}
 \left[
\begin{array}{cccc}
b_o&&&\\
& b_1&0&\\
&& \ddots & \\
& 0 &&b_{N-1}
\end{array}
\right]
\]
i.e. $\mathbf{W}$ is obtained by pre- and post multiplying
$\mathbf{C}$ by two diagonal matrices. It is easy to see that we
have
\[
\wm \!=\mathbf{C}_{[c_0, c_1, \ldots, c_{N-1}]}\odot 
\left[
\begin{array}{llll}
a_0b_0& a_0b_1& \ldots&a_0b_{N-1}\\
a_1b_0& a_1b_1& \ldots&a_1b_{N-1}\\
 \vdots& \vdots&&\vdots\\
a_{N-1}b_0 & a_{N-1}b_1& \ldots & a_{N-1}b_{N-1}
 \end{array} 
\!\!\right]
\!=\mathbf{C}\odot \mm
\]
where $\odot$ stands for the  Schur Hadamard  multiplication (or a
``masking" operation) which multiplies matrices element-wise, and 
\[
\mm\triangleq [a_kb_l]_{k,l=0,\ldots,n-1}.
\]
For matrices of the type $\mathbf{W}$, we have that they inherit 
interesting diagonalization properties from the original circulant 
$\mathbf{C}$. The matrix $\wm$ is a circulant matrix that is modified 
by a highly structured masking matrix $ \mm$ and we have that
\[
\wm=\textrm{Diag}[a_0,\ldots, a_{N-1}][\mathbf{FT}] 
\textrm{Diag}[\mu_0,\ldots, \mu_{N-1}][\mathbf{FT}]^*
\textrm{Diag}[b_0,\ldots, b_{N-1}].
\]
However, since the masking matrix is neither circulant nor Toeplitz, 
we shall have to consider some special cases for the 
$\{a_0, a_1, \ldots, a_{N-1}\}$ and $\{b_0, b_1, \ldots, b_{N-1}\}$ 
sequences. First of all, note that the factorization above will be of
the form
\[
\wm=\mathbf{U}\left[
\begin{array}{ccc}
\mu_0&&\\
& \ddots&\\
&& \mu_{N-1}
\end{array}
\right] \mathbf{U}^{-1}
\]
if and only if
\begin{eqnarray*}
(\textrm{Diag}[a_0, a_1, \ldots, a_{N-1}][\mathbf{FT}])^{-1} &=&
[\mathbf{FT}]^*\textrm{Diag}[b_0, b_1, \ldots, b_{N-1}] \\
\iff [\mathbf{FT}]^*\textrm{Diag}[a_0^{-1}, a_1^{-1}, \ldots, a_{N-1}^{-1}]&=&
[\mathbf{FT}]^*\textrm{Diag}[b_0, b_1, \ldots, b_{N-1}] 
\end{eqnarray*}
or $b_k=a^{-1}_k$, and $\mathbf{U}$ will further be unitary if also
$b_k=a^*_k$, implying that $a_k=e^{j\alpha_k}$ and 
$b_k=e^{-j\alpha_k}=a_k^*$. In this case the masking-matrix
multiplying $\mathbf{C}$ will be
$[e^{j\alpha_k}e^{-j\alpha_l}]=[e^{j(\alpha_k-\alpha_l)}]_{k,l=0,\ldots,N-1}$. 

The most interesting particular cases of $\{a_0, a_1, \ldots, a_{N-1}\}$ and
$\{b_0, b_1, \ldots, b_{N-1}\}$ arise when we have $a_k=\gamma^k$
and $b_k=\gamma^{-k}$, $k=0,1\ldots,N-1$, for some real or imaginary $\gamma$. 
In this case, we have in general
\begin{eqnarray*}
\mm&=&
\left[
\begin{array}{lllll}
1& \gamma^{-1}& \gamma^{-2}& \ldots& \gamma^{-(N-1)}\\
\gamma & 1 & \gamma^{-1}&\ldots& \gamma^{-(N-1)+1}\\
\gamma^2 & \gamma &1&\ldots& \gamma^{-(N-1)+2}\\
\vdots& \vdots& \ddots & \ddots &\vdots \\
\gamma^{N-1}& \gamma^{N-2}& \ldots& \gamma& 1
\end{array}
\right]
\\
&=&
\textrm{Circ}_{[1, \gamma^{-1}, \ldots, \gamma^{-(N-1)}]}
 \odot
\left[
\begin{array}{ccccc}
1&1&1&1&1\\
\gamma^N&1&1&1&1\\
\gamma^N&\gamma^N&1&1&1\\
\vdots&&&\ddots&\\
 \gamma^N&\gamma^N&\ldots&\gamma^N&1
\end{array}
\right]
\end{eqnarray*}
where $\textrm{Circ}_{[1, \gamma^{-1}, \ldots, \gamma^{-(N-1)}]}$ is 
given by
\[
\left[
\begin{array}{ccccc}
1& \gamma^{-1}& \gamma^{-2}& \ldots& \gamma^{-(N-1)}\\
\gamma^{-(N-1)} & 1 & \gamma^{-1}&\ldots& \gamma^{-(N-1)+1}\\
\gamma^{-(N-1)+1} & \gamma^{-(N-1)} &1&\ldots& \\
\vdots& \vdots& \ddots & 1 &\vdots \\
\gamma^{-(N-1)+(N-2)}& & \ldots& \gamma^{-(N-1)}& 1
\end{array}
\right].
\]
Hence the matrix $\wm=\mathbf{C}\odot \mm$ becomes
\[
\wm=\mathbf{C}_{[c_0, \ldots, c_{N-1}]} \odot \textrm{Circ}_{[1,
\gamma^{-1}, \ldots, \gamma^{-(N-1)}]}\odot \left[
\begin{array}{ccccc}
1&1&1&1&1\\
\gamma^N&1&1&1&1\\
\gamma^N&\gamma^N&1&1&1\\
\vdots&&&\ddots&\\
 \gamma^N&\gamma^N&\ldots&\gamma^N&1
\end{array}
\right]
\]
which clearly is a $\lambda ( =\gamma^N)$- circulant matrix.

To summarize, we have the following result: A $\lambda$-circulant 
matrix $\wm$, denoted by
\[
\wm=\left[
\begin{array}{ccccc}
m_0 & m_1& m_2 & \ldots& m_{N-1}\\
\lambda m_{N-1}& m_0& m_1& \ldots& m_{N-2}\\
\lambda m_{N-2}& \lambda m_{N-1}& m_0&\ldots& \ldots\\
\vdots & \vdots & \ldots &\ddots& \ldots\\
\lambda m_1 & \lambda m_2 & \ldots & \lambda m_{N-1}& m_0
\end{array}
\right]
\]
can be rewritten as
\[
\wm=\textrm{Circ}_{[m_0, m_1\gamma, m_2\gamma^2, \ldots,
m_{N-1}\gamma^{N-1}]}\odot \textrm{Circ}_{[1, \gamma^{-1}, \ldots,
\gamma^{-(N-1)}]}\odot \Lambda
\]
with
\[
\Lambda=\left[
\begin{array}{ccccc}
1&1&1&\ldots&1\\
\lambda&1&1&\ldots&1\\
\lambda&\lambda&1&\ldots&1\\
\vdots& \vdots& \lambda& 1&\vdots\\
\lambda&\lambda& \ldots&\lambda&1
\end{array}
\right] \textrm{and  } \gamma^N=\lambda
\]
and hence can be factorized as 
\[
\wm=
\left[
\begin{array}{l}
1\\
~~~\gamma~~~~ \\
~~~~~~\gamma^2~~ \\
~~~~~~~~~\ddots~~ \\
~~~~~~~~~~~~\gamma^{N-1}
\end{array}\right]
[\mathbf{FT}]
\left[
\begin{array}{ccccc}
\mu_0&&&&\\
& \mu_1&&0&\\
&&\mu_2&& \\
&0&&\ddots & \\
& & &&\mu_{N-1}
\end{array}\right]
[\mathbf{FT}]^*
\left[
\begin{array}{l}
1\\
~~~\gamma^{-1}\\
~~~~~~\gamma^{-2} \\
~~~~~~~~~\ddots~~ \\
~~~~~~~~~~~~ \gamma^{-(N-1)}
\end{array}\right]
\]
where $[\mu_0, \mu_1, \ldots, \mu_{N-1}]$ are the eigenvalues of
\[
\textrm{Circ}_{[m_0, m_1\gamma,  \ldots, m_{N-1}\gamma^{N-1}]}
\triangleq \textrm{Circ}_{[c_0, c_1, \ldots, c_{N-1}]}
\]
given by
\[
\mu_l=\sum^{N-1}_{k=0} m_k\cdot \gamma^k\cdot 
e^{-i\frac{2\pi}{N}kl}~~~(\gamma\triangleq\lambda^{\frac{1}{N}}).
\]
Therefore $\wm$ is readily diagonalized as follows
\begin{eqnarray*}
\left[
\begin{array}{ccc}
\mu_0&&\\
&\ddots&\\
&&\mu_{N-1}
\end{array}\right]
\!\!\!\!\!
&=& 
[\mathbf{FT}]^*
\left[
\begin{array}{l}
1\\
~~~\gamma^{-1}\\
~~~~~~\gamma^{-2} \\
~~~~~~~~~\ddots~~ \\
~~~~~~~~~~~~ \gamma^{-(N-1)}
\end{array}\right]
\wm
\left[
\begin{array}{l}
1\\
~~~\gamma\\
~~~~~~\gamma^2 \\
~~~~~~~~~\ddots~~ \\
~~~~~~~~~~~~ \gamma^{N-1}
\end{array}\right]
[\mathbf{FT}]\\
&=&
\tm^{-1}
\wm \tm,
\end{eqnarray*}
the matrices $\tm$ and $\tm^{-1}$ being
\[
\tm=\left[\begin{array}{cccc}
1&&&\\
&\gamma&&\\
&&\ddots&\\
&&&\gamma^{N-1}
\end{array}\right]
[\mathbf{FT}] \textrm{   and   }
\tm^{-1}= [\mathbf{FT}]^*
\left[\begin{array}{cccc}
1&&&\\
&\gamma^{-1}&&\\
&&\ddots&\\
&&&\gamma^{N-1}
\end{array}\right]. 
\]
Note that $\tm$ is not, in general a unitary transformation. In all 
developments above, we assumed $\gamma$ to be arbitrary. 
If $\gamma\neq 0$ is a real number, $\tm$ will be an invertible matrix, 
as seen before. If however $\gamma$ is purely 
imaginary, i.e. $\gamma=e^{j\varphi}$, then clearly
$\gamma^*=e^{-j\varphi}=\gamma^{-1}$ and the matrix $\tm$ becomes
a unitary transformation, obeying
\[
\tm\tm^*=\tm^*\tm=I.
\]
In this case the matrix $\wm$ will be $\lambda$-factor circulant with
$\lambda=e^{j\varphi N}$.

%
%

\section{Dynamics of a Cyclically Interacting Swarm}

Returning to the problem of analyzing the dynamics and the long-term
behavior of a swarm of robots $\mathcal{P}_0, \mathcal{P}_1, \ldots,
\mathcal{P}_{N-1}$ interacting according to
\begin{eqnarray*}
\left.
\begin{array}{ll}
 & \Pm(t+1)\\
 \textrm{or} & \frac{d}{dt}\Pm(t)
 \end{array}
\right\}
&=&\left[\begin{array}{ccccl}
m_0 & m_1& m_2 & \ldots& m_{N-1}\\
\lambda m_{N-1}& m_0& m_1& \ldots& m_{N-2}\\
\lambda m_{N-2}& \lambda m_{N-1}& m_0&\ldots& \ldots\\
\vdots & \vdots & \ldots &\ddots& \ldots\\
\lambda m_1 & \lambda m_2 & \ldots & \lambda m_{N-1}& m_0
\end{array}
\right]\Pm(t)\\
&&\\
&=&\Phi\Pm(t),
\end{eqnarray*}
we have that the interaction matrix $\Phi$ is $\lambda$-circulant
hence it is diagonalizable as follows:
\[
\Phi = 
\left[
\begin{array}{l}
1\\
~~~\gamma\\
~~~~~~\gamma^2 \\
~~~~~~~~~\ddots~~ \\
~~~~~~~~~~~~ \gamma^{N-1}
\end{array}\right]
[\mathbf{FT}] 
\left[
\begin{array}{cccc}
\mu_0&&&\\
&\ddots&&0\\
&0&\ddots&\\
&&&\mu_{N-1}
\end{array}\right]
[\mathbf{FT}]^* 
\left[
\begin{array}{l}
1\\
~~~\gamma^{-1}\\
~~~~~~\gamma^{-2} \\
~~~~~~~~~\ddots~~ \\
~~~~~~~~~~~~ \gamma^{-(N-1)}
\end{array}\right]
\]
where $\gamma=\lambda^{\frac{1}{N}}$ and
\[
\mu_l=\sum_{k=0}^{N-1}m_k\lambda^{\frac{k}{N}}e^{-i\frac{2\pi}{N}kl}.
\]
Therefore defining
\[
\mathbf{\widetilde{P}}(t)
\triangleq 
[\mathbf{FT}]^* 
\left[
\begin{array}{cccc}
1&&&\\
&\lambda^{-\frac{1}{N}}&&\\
&&\ddots&\\
&&&\lambda^{-\frac{N-1}{N}}
\end{array}
\right]\Pm(t)
\]
we have decoupled dynamics for the transformed location vector, given by
\[
\left.
\begin{array}{l}
\frac{d}{dt}\widetilde{\Pm}(t)\\
\textrm{      or} \\
\mathbf{\widetilde{P}}(t+1)
\end{array}
\right\}= 
\left[
\begin{array}{cccc}
\mu_0&&&\\
&\mu_1&0&\\
&0&\ddots&\\
&&&\mu_{N-1}
\end{array}\right]\widetilde{\Pm}(t)
\]
and the evolution of the swarm is controlled by the eigenvalues
$\mu_0, \mu_1, \ldots, \mu_{N-1}$.

Let us concentrate next on some specific cases of 
$\overline{m}=[m_0,\ldots,m_{N-1}]$ and $\lambda$. A ``$\lambda$- cyclic" 
interaction involves agents that are reacting differently with 
the agents that follow them to the agents that precede them in the 
ordering $\Pc_0,\ldots,\Pc_{N-1}$.


\subsection{Darboux's polygon evolution and extensions}

As a first example, suppose that we have a generalization of
Darboux's polygon evolution process \cite{D}, which is also a nice model 
for cyclic pursuit:
\[
\Pm(t+1)=\left[
\begin{array}{cccccc}
\frac{1}{2}&\frac{1}{2}&0&0&\ldots&\\
0&\frac{1}{2}&&&&\\
0&&\ddots&&&\\
0&&&\ddots&&\\
0&&&&\frac{1}{2}&\frac{1}{2}\\
\lambda\frac{1}{2}&0&0&0&0&\frac{1}{2}
\end{array}
\right] \Pm(t).
\]
In this case, we have a $\lambda$-factor circulant with
\[
\mu_l
=\frac{1}{2}+\frac{1}{2}\lambda^{\frac{1}{N}}e^{-i\frac{2\pi}{N}\cdot l} 
=\frac{1}{2}(1+\lambda^{\frac{1}{N}}e^{-i\frac{2\pi}{N}\cdot l}),
~l=0,1,\ldots,N-1.
\]
Here, the evolution of the polygon vertices (or the agents in cyclic 
pursuit) is described by
\[
\mathbf{\widetilde{P}}(t+1)=
\left[
\begin{array}{cccc}
\mu_0^t&&&\\
&\mu_1^t&0&\\
&&\ddots&\\
&0&&\mu_{N-1}^t
\end{array}
\right]\mathbf{\widetilde{P}}(0)
\]
where we defined
\[
\mathbf{\widetilde{P}}(t)
= [\mathbf{FT}]^* 
\left[
\begin{array}{cccc}
1&&&\\
&\lambda^{-1/N}&0&\\
&&\ddots&\\
&0&& \lambda^{-(N-1)/N}
\end{array}
\right]\Pm(t).
\]
From this we have
\begin{eqnarray*}
\Pm(t)
&=&
\left[
\begin{array}{cccc}
1&&&\\
& \lambda^{1/N}&&\\
&&\ddots&\\
&&&\lambda^{\frac{N-1}{N}}
\end{array}
\right]
[\mathbf{FT}]\widetilde{\Pm}(t)\\
&=&
\left[
\begin{array}{cccc}
1&&&\\
& \lambda^{1/N}&&\\
&&\ddots&\\
&&&\lambda^{\frac{N-1}{N}}
\end{array}
\right]
[\mathbf{FT}]
\left[
\begin{array}{cccc}
\mu_0^t&&&\\
&\mu_1^t&0&\\
&0&\ddots&\\
&&&\mu_{N-1}^t
\end{array}\right]
\widetilde{\Pm}(0).
\end{eqnarray*}

The evolution of the polygon vertices (the swarm of robots) when we let 
the time grow, thus asymptotically depends on the dominant eigenvalues 
among $\mu_0,\ldots,\mu_{N-1}$.

\begin{figure}
\hspace{-1cm}
\includegraphics[scale=0.4]{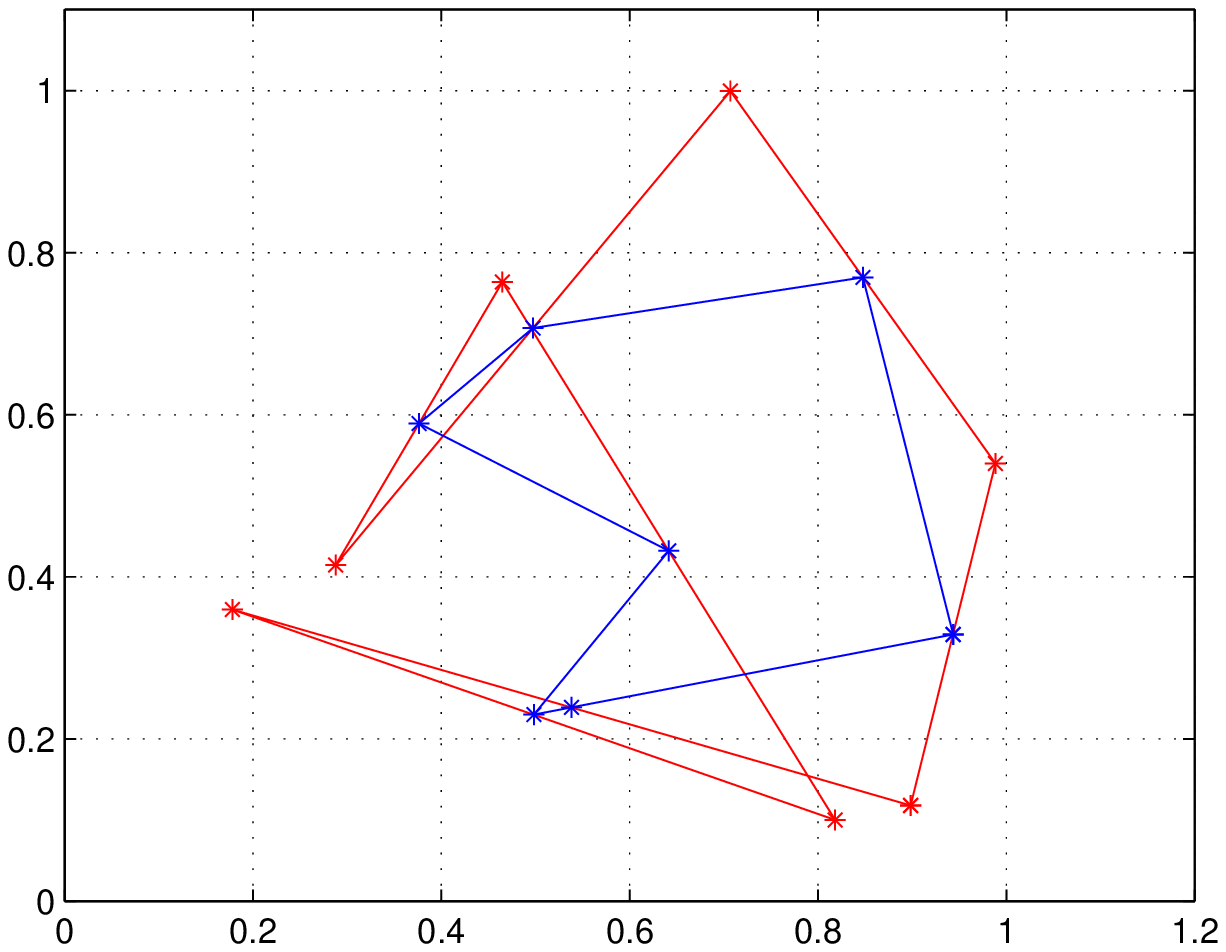}\includegraphics[scale=0.4]{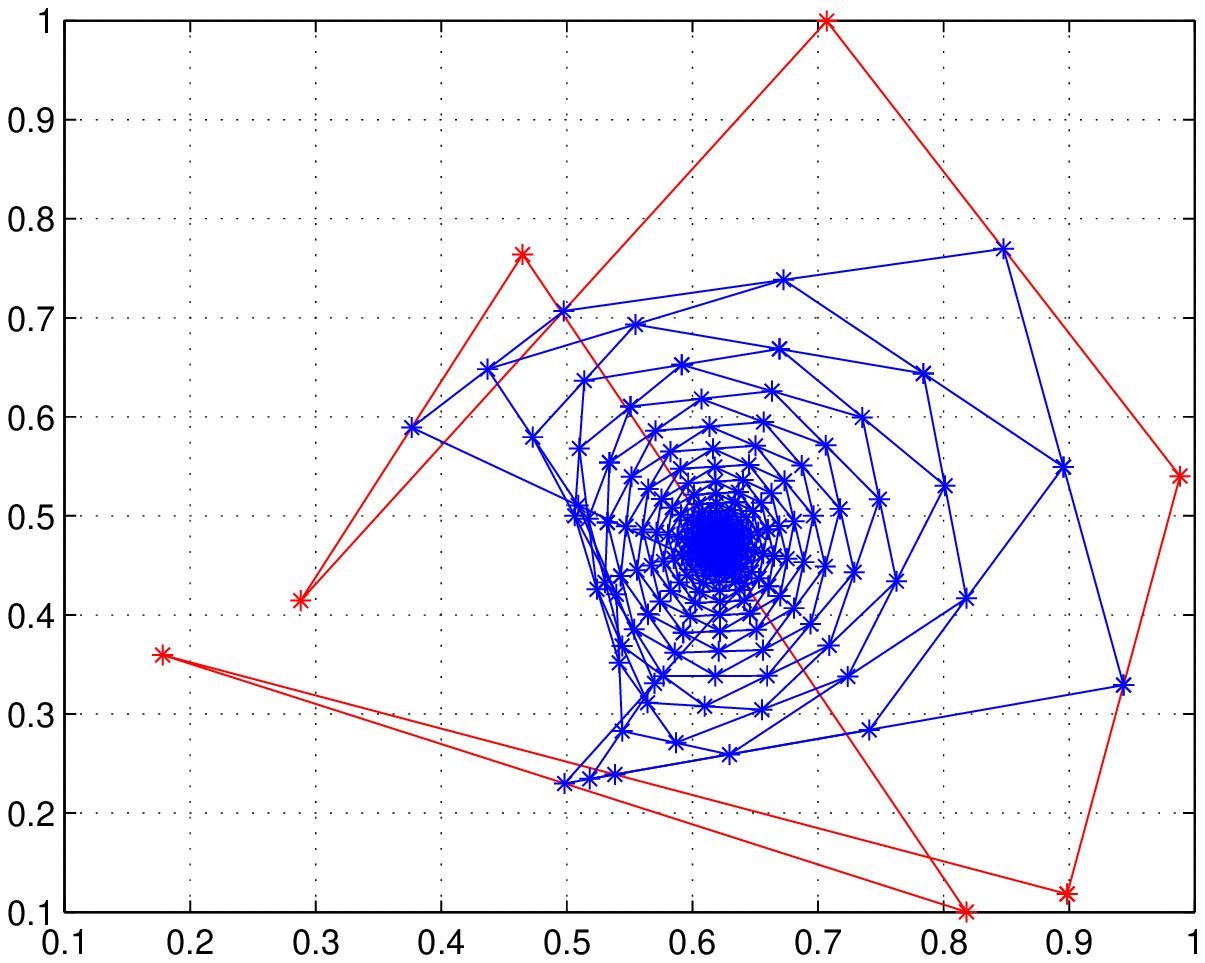}\includegraphics[scale=0.4]{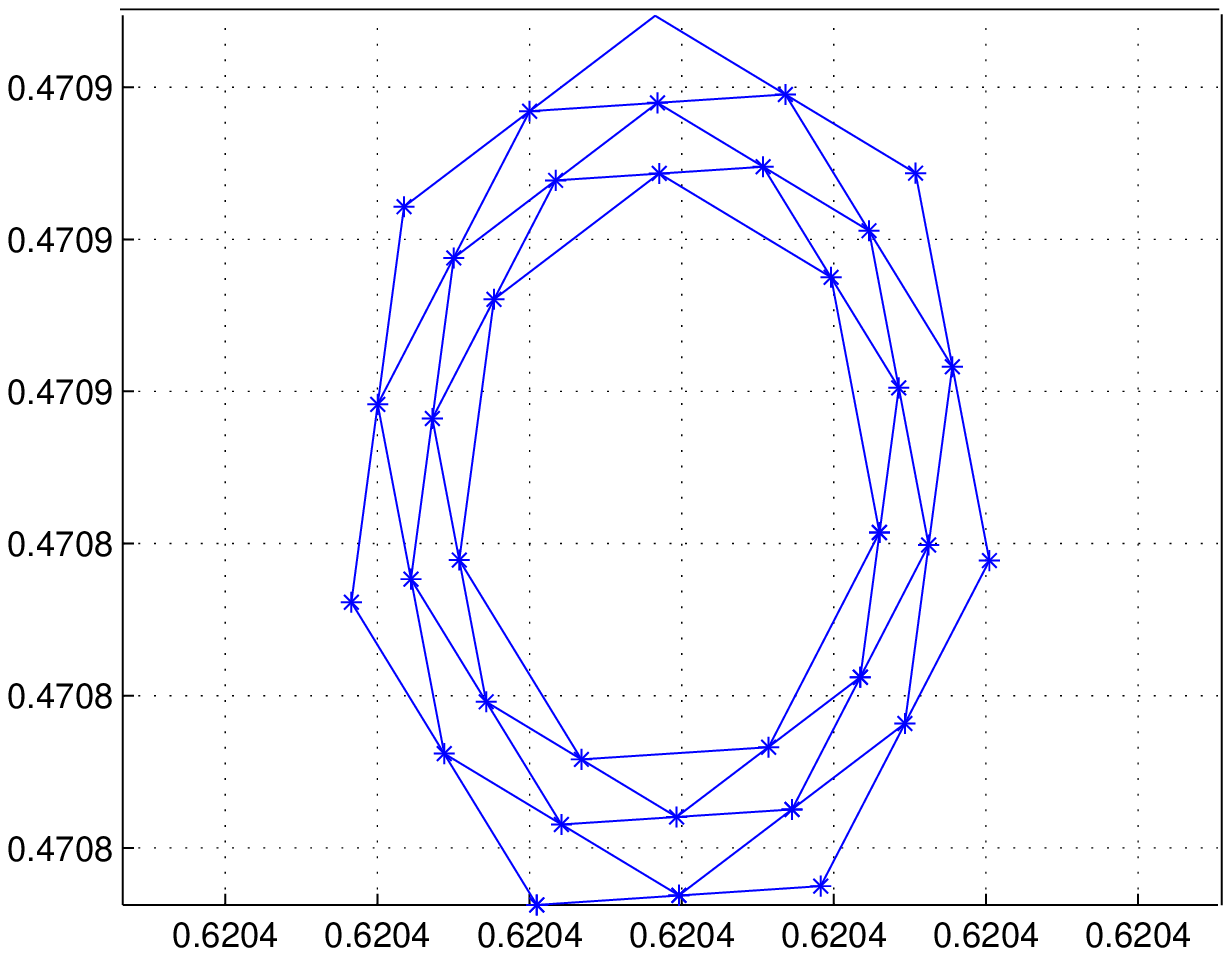}
\caption{The cyclic pursuit case ($\lambda=1$) with a random initial polygon 
with $N=7$ points, the first figure presents 
the initial configuration (in red) and the first iteration (in blue), the 
second shows the entire evolution for 100 iterations, the last figure displays 
the scaled up configuration for the last few iterations.}
\label{fig:gamma1}
\end{figure}

In the case of $\lambda=1$ (or circulant cyclic pursuit), 
we have 
\[
\mu_l=\frac{1}{2}(1+e^{-i\frac{2\pi}{N}\cdot l}),~l=0,1,\ldots,N-1,
\]
and $\mu_0=1$. Then
\begin{eqnarray*}
\Pm(t)_{t\rightarrow\infty}
&=& 
[\mathbf{FT}]
\left[
\begin{array}{cccc}
\mu_0^t&&&\\
&\mu_1^t&0&\\
&0&\ddots&\\
&&&\mu_{N-1}^t
\end{array}\right]_{t\rightarrow\infty}
\!\!\!\!\!\widetilde{\Pm}(0)\\
&=&
[\mathbf{FT}]
\left[
\begin{array}{cccccc}
1&&&&&\\
& \mu_1^t&0& &&\\
& &0 &  && \\
&0&&\ddots &&\\
&&&&0&\\
&&&&& \mu_{N-1}^t
\end{array}\right]_{t\rightarrow\infty}
\!\!\!\!\!
[\mathbf{FT}]^*
\Pm(0).
\end{eqnarray*}
Since the dominant eigenvalue $\mu_0=1$ and all others have modulus 
less than one, we have that the limiting behavior is
\[
\Pm(t)_{t\rightarrow\infty}
=
\frac{1}{N}
\left[
\begin{array}{c}
1\\
1\\
\vdots\\
1
\end{array}
\right]
[
1,1,\ldots,1
]\Pm(0).
\]
Hence the point constellation converges to the centroid of the
initial locations. The way this convergence occurs is be controlled 
by the next dominant eigenvalues, which are in this case
\begin{eqnarray*}
\mu_1&=&\frac{1}{2}(1+e^{-i\frac{2\pi}{N}})\\
\mu_{N-1}&=&\frac{1}{2}(1+e^{-i\frac{2\pi(N-1)}{N}}).
\end{eqnarray*}
Indeed, writing
\[
\Pm^N(t)
=
\Pm(t)-
\frac{1}{N}
\left[
\begin{array}{c}
1\\
1\\
\vdots\\
1
\end{array}
\right]
[
1,1,\ldots,1
]\Pm(0),
\]
we have
\[
\Pm^N(t)=
[\mathbf{FT}]
\left[
\begin{array}{cccc}
0&&&\\
&\mu_1&0&\\
&0&\ddots&\\
&&&\mu_{N-1}
\end{array}
\right]
[\mathbf{FT}]^*\Pm(0)
\]
and, disregarding the faster decaying terms $\mu_i^t$, $i=2,\ldots,N-2$,  
we further get
\begin{eqnarray*}
\Pm^N(t)_{t\rightarrow\infty}
&=&
\frac{1}{N}
\left[
\begin{array}{c}
1\\
w\\
\vdots\\
w^{N-1}
\end{array}
\right]
[
1,w,\ldots,w^{N-1}
]\Pm(0)\mu_1^t\\
&&+
\frac{1}{N}
\left[
\begin{array}{c}
1\\
w^{N-1}\\
\vdots\\
w^{(N-1)(N-1)}
\end{array}
\right]
[
1,w^{N-1},\ldots,w^{(N-1)(N-1)}
]\Pm(0)\mu_{N-1}^t.
\end{eqnarray*}
Hence
\[
\Pm^N(t)_{t\rightarrow\infty}
=
\frac{1}{N}
\left[
\begin{array}{c}
1\\
w\\
\vdots\\
w^{N-1}
\end{array}
\right]
A(t)\mu_1^t+
\frac{1}{N}
\left[
\begin{array}{c}
1\\
w^{N-1}\\
\vdots\\
w^{(N-1)(N-1)}
\end{array}
\right]B(t)\mu_{N-1}^t
\]
where $A(t)\mu_1^t$ and $B(t)\mu_{N-1}$ are some complex numbers, and 
$\Pm^N(t)$ will be, in the limit $t\rightarrow\infty$, an affine 
transformation of a regular polygon, i.e. a discrete ellipse (see Figure 
\ref{fig:gamma1}).

\begin{figure}
\hspace{-1cm}
\includegraphics[scale=0.4]{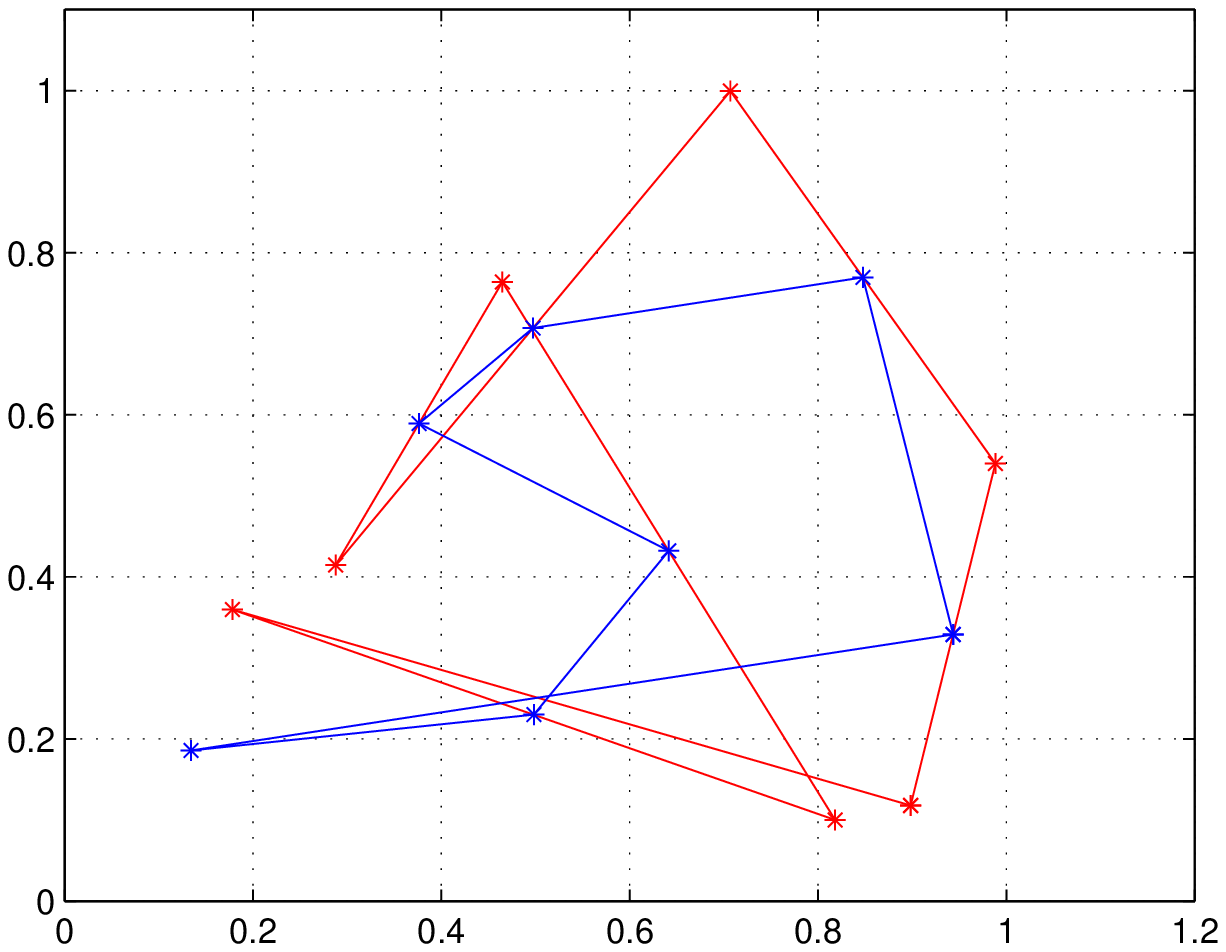}\includegraphics[scale=0.4]{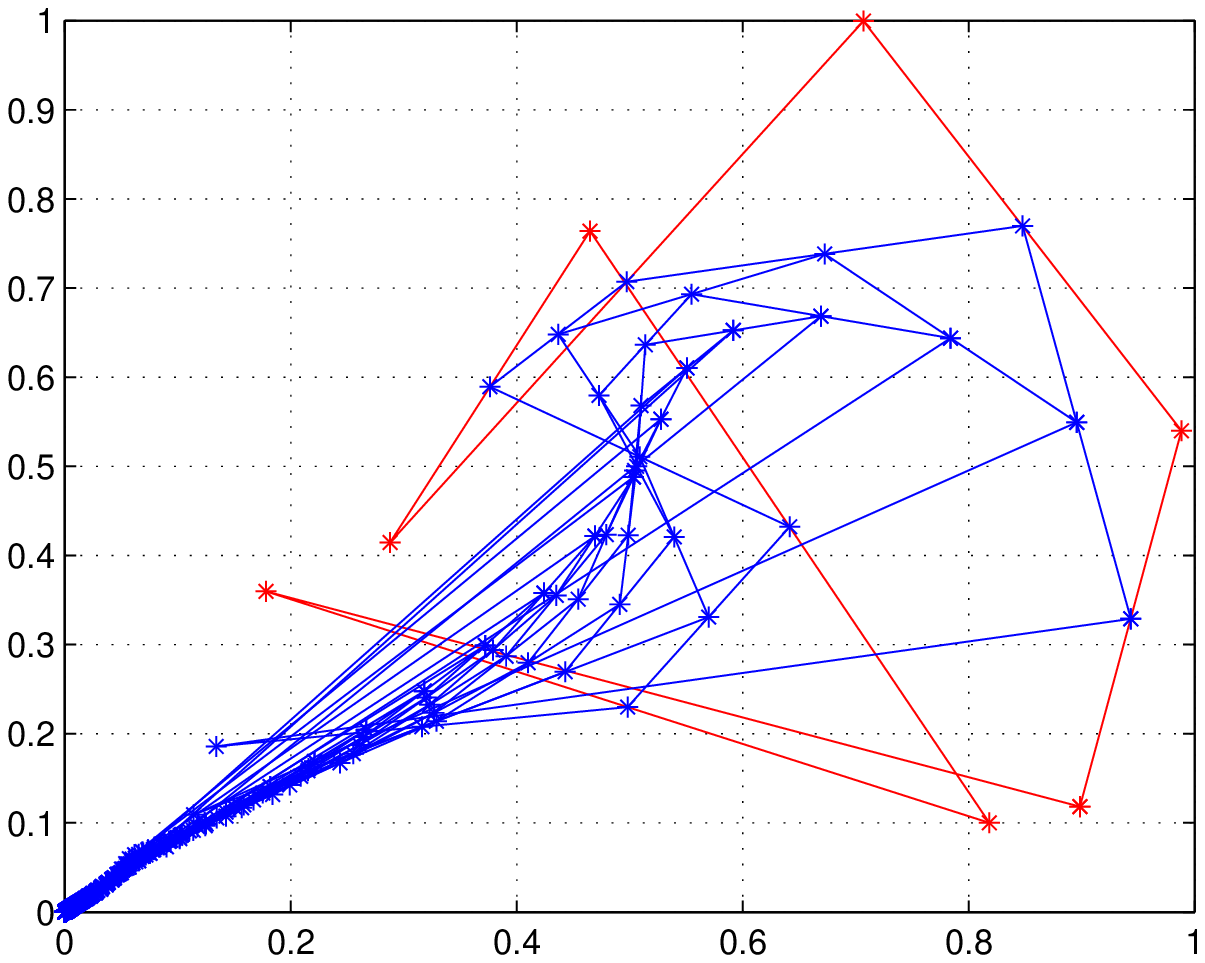}\includegraphics[scale=0.4]{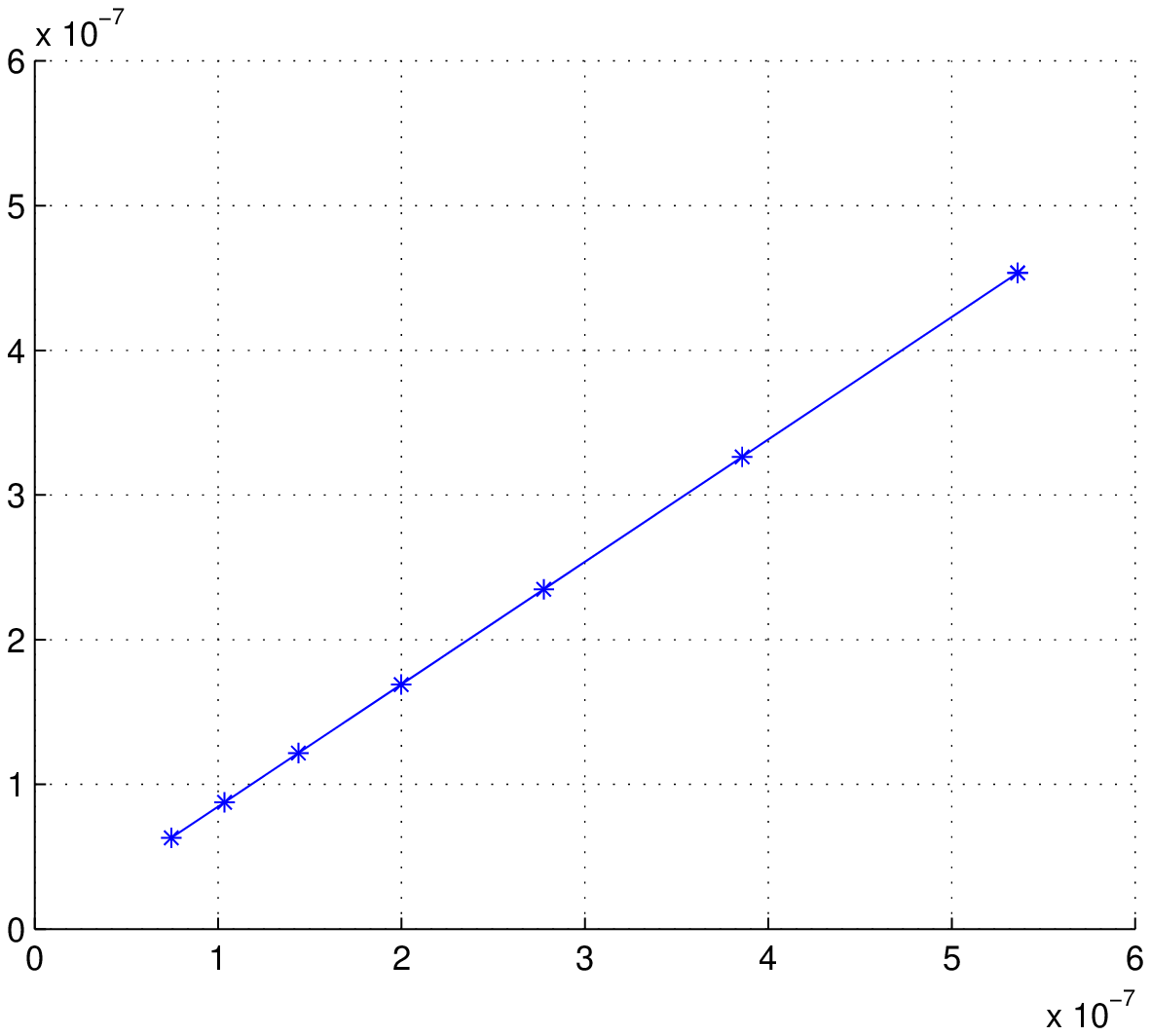}
\caption{$\lambda=0.1$}
\label{fig:gamma01}
\end{figure}

\begin{figure}
\hspace{-1cm}
\includegraphics[scale=0.4]{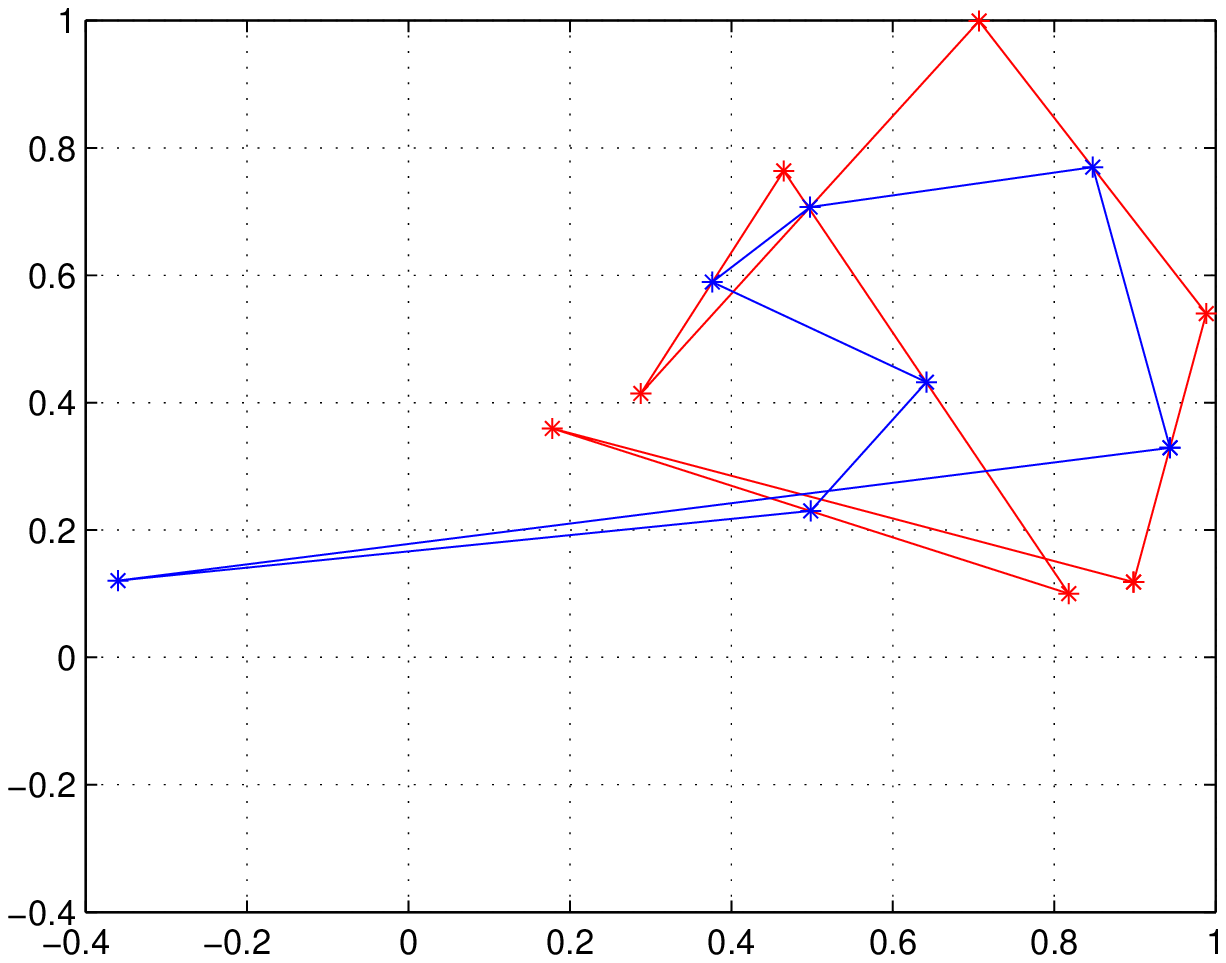}\includegraphics[scale=0.4]{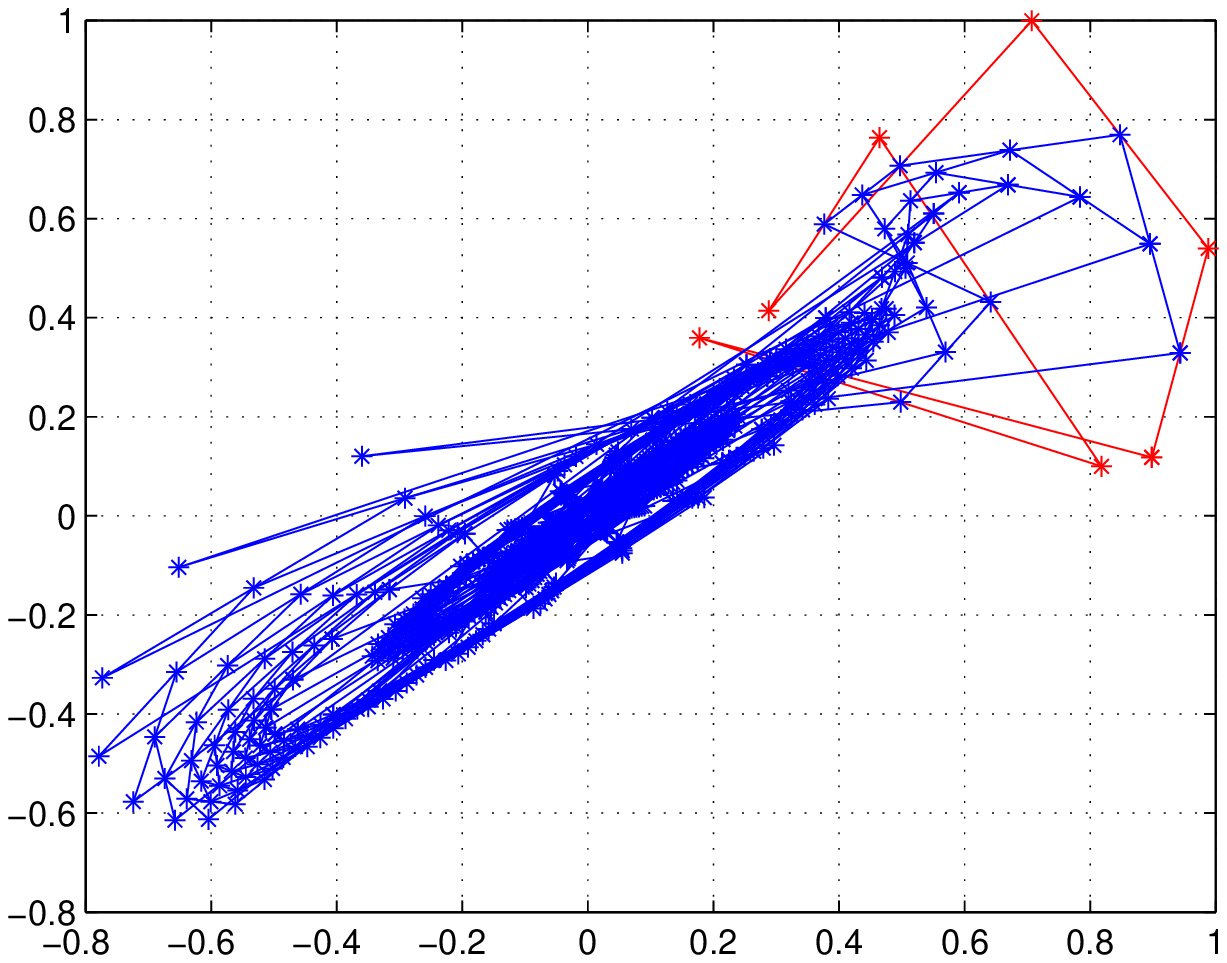}\includegraphics[scale=0.4]{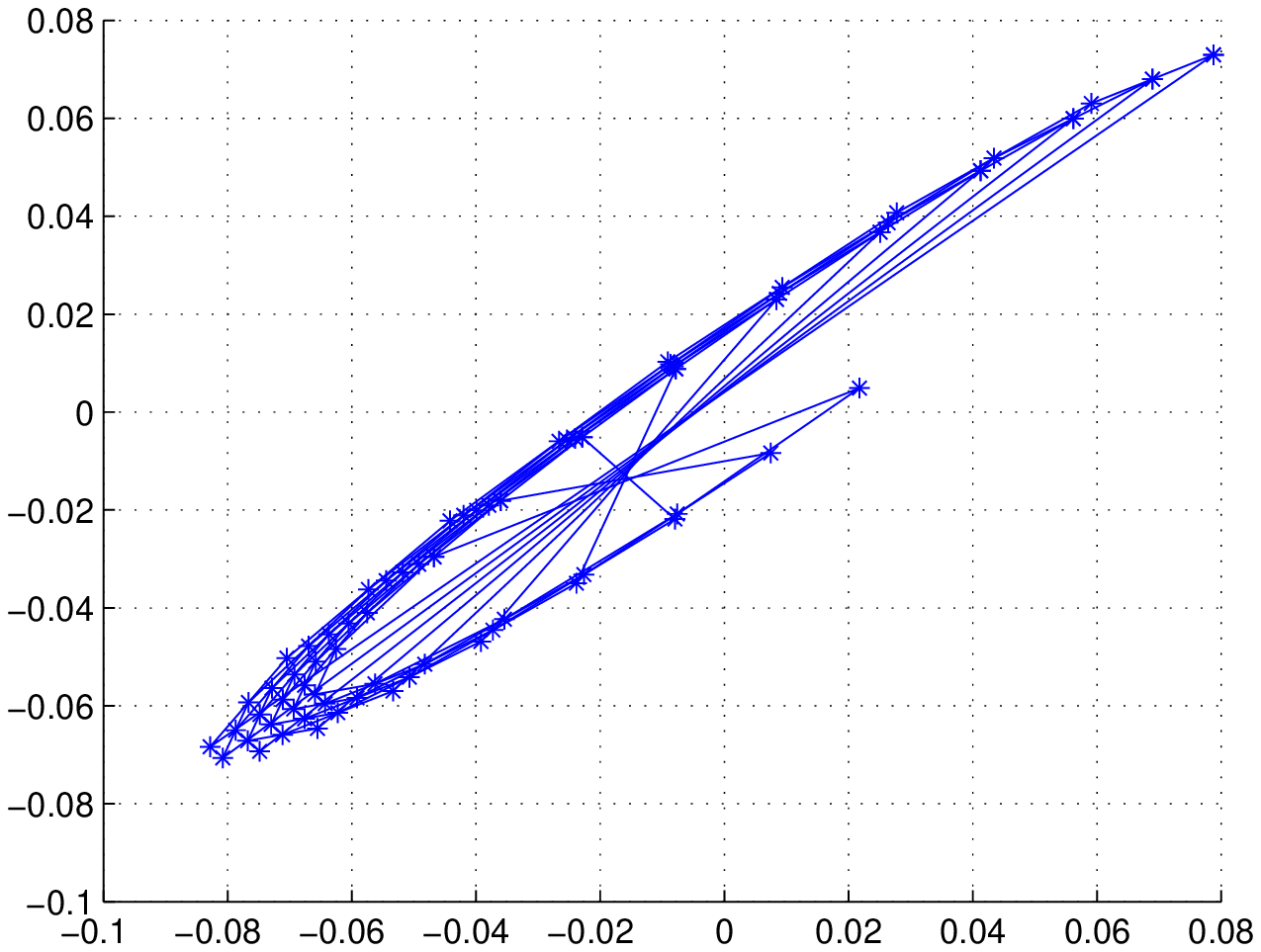}
\caption{$\lambda=-1$}
\label{fig:gamma_1}
\end{figure}

\begin{figure}
\hspace{-1cm}
\includegraphics[scale=0.4]{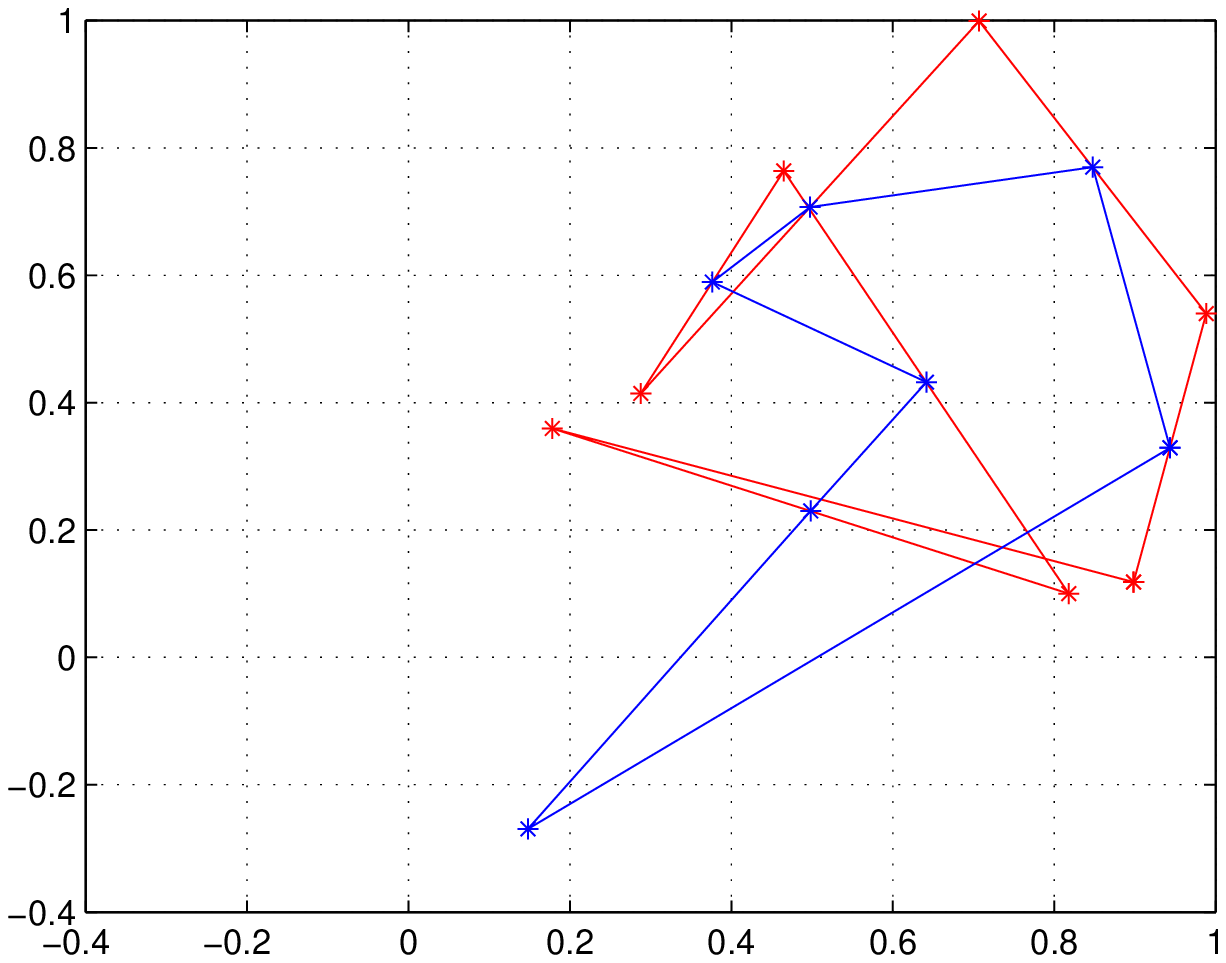}\includegraphics[scale=0.4]{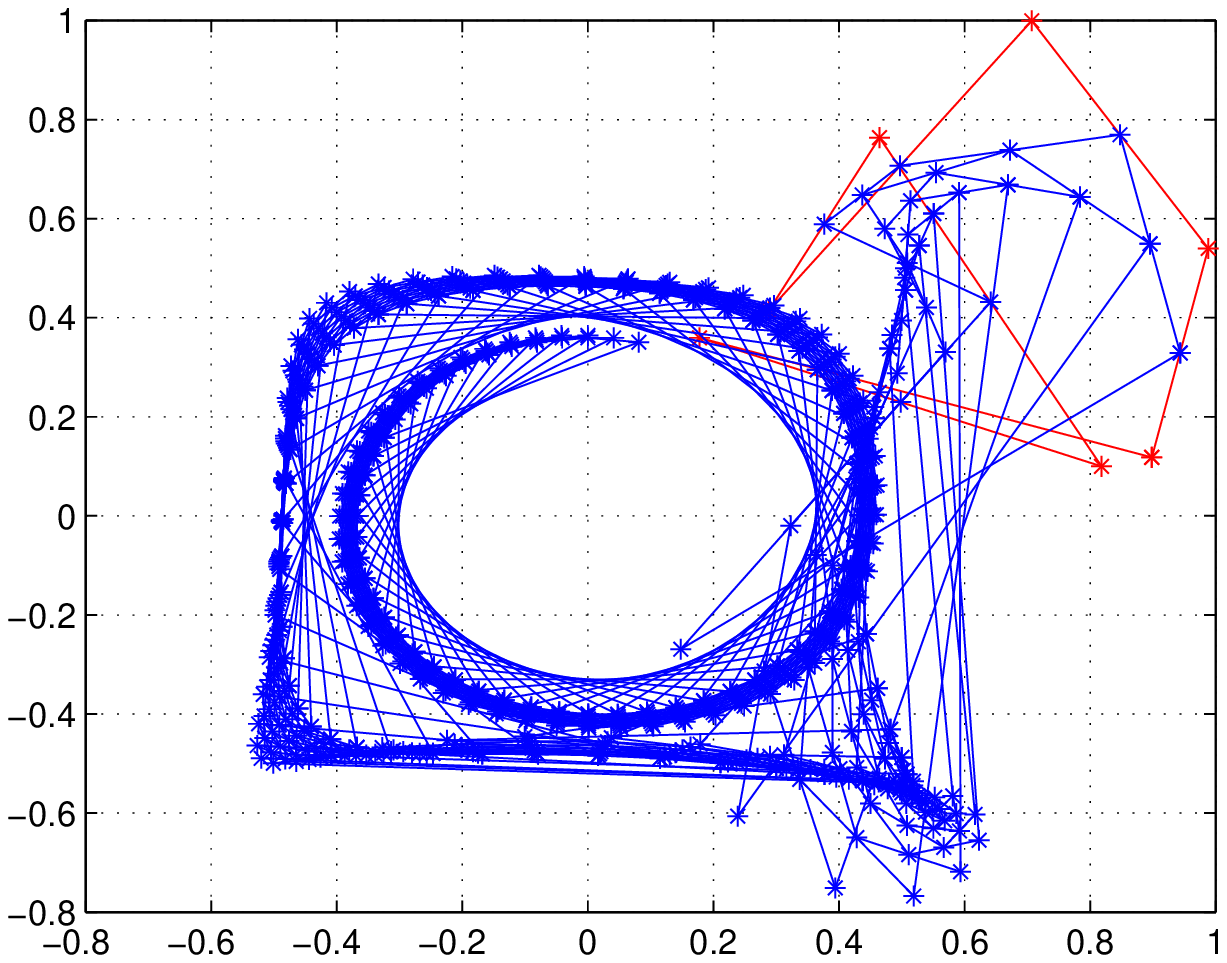}\includegraphics[scale=0.4]{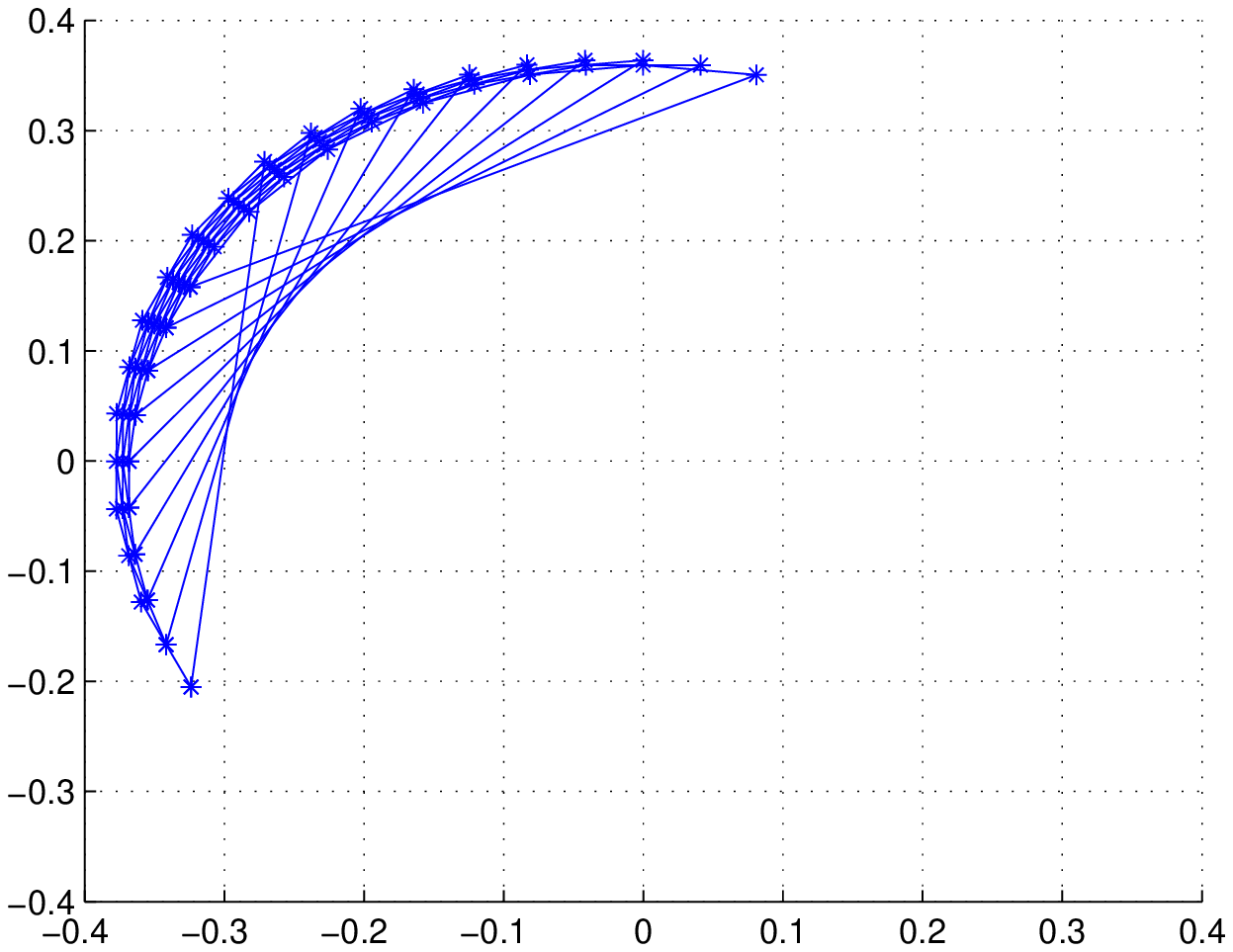}
\caption{$\lambda=i$}
\label{fig:gammai}
\end{figure}

\begin{figure}
\hspace{-1cm}
\includegraphics[scale=0.4]{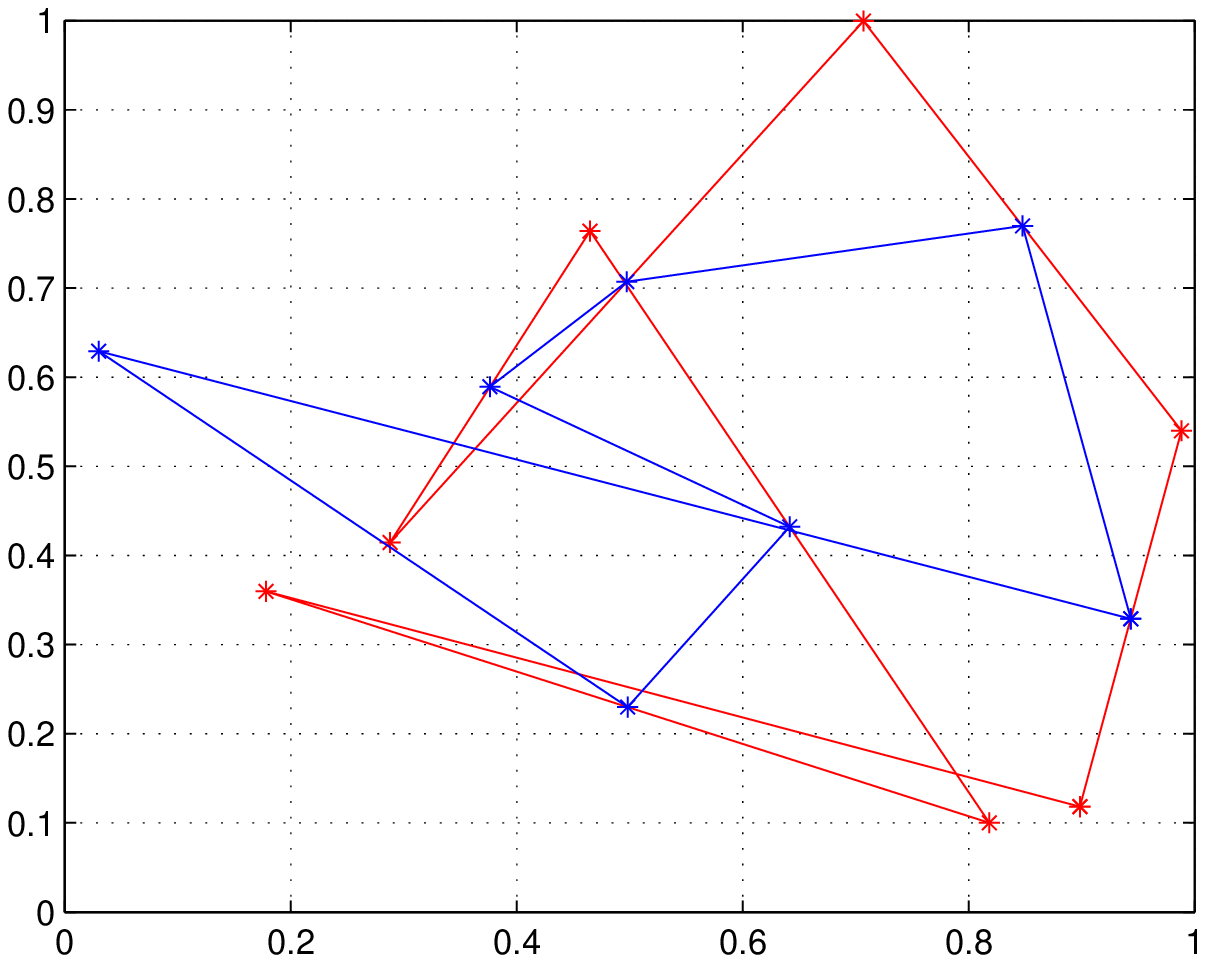}\includegraphics[scale=0.4]{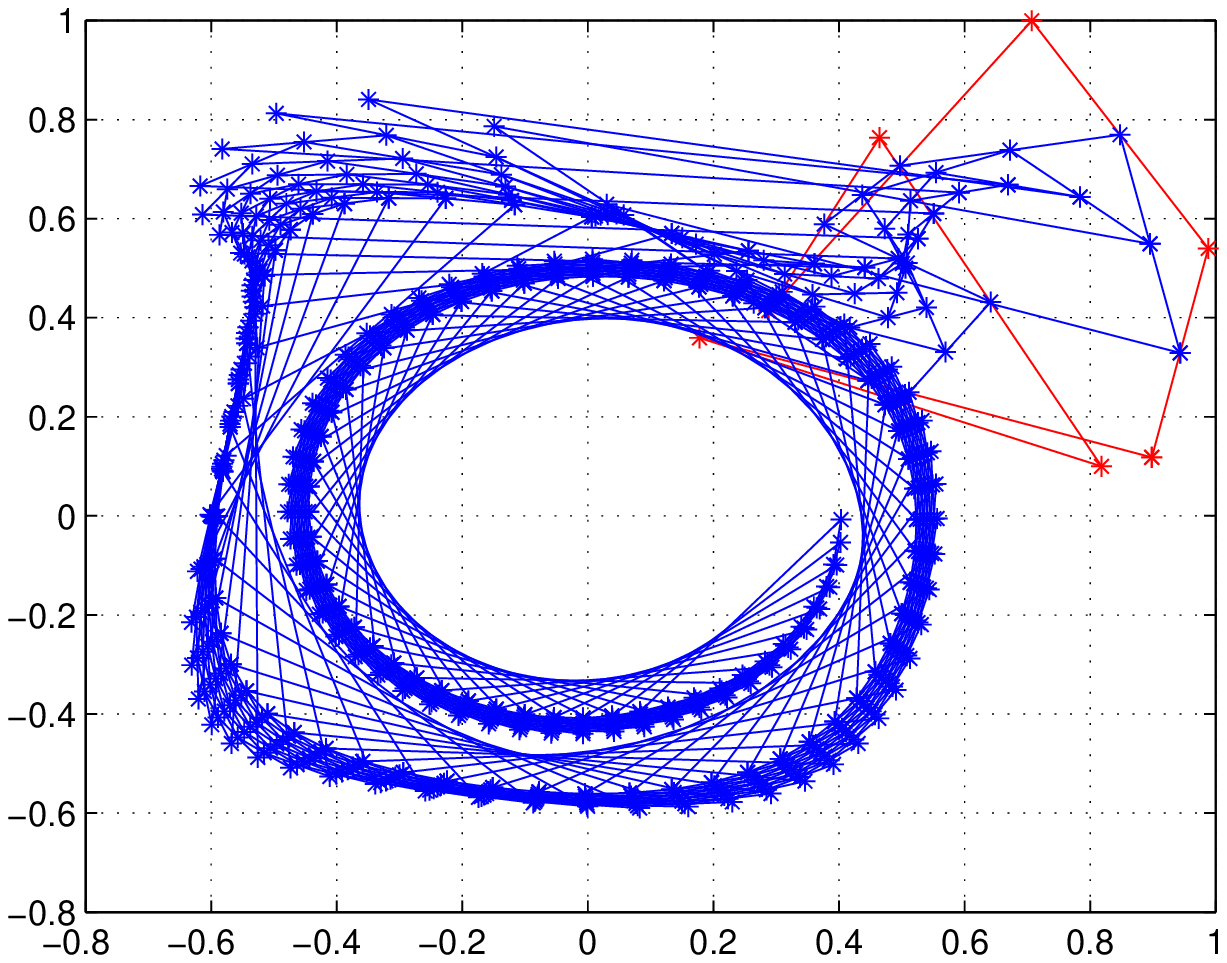}\includegraphics[scale=0.4]{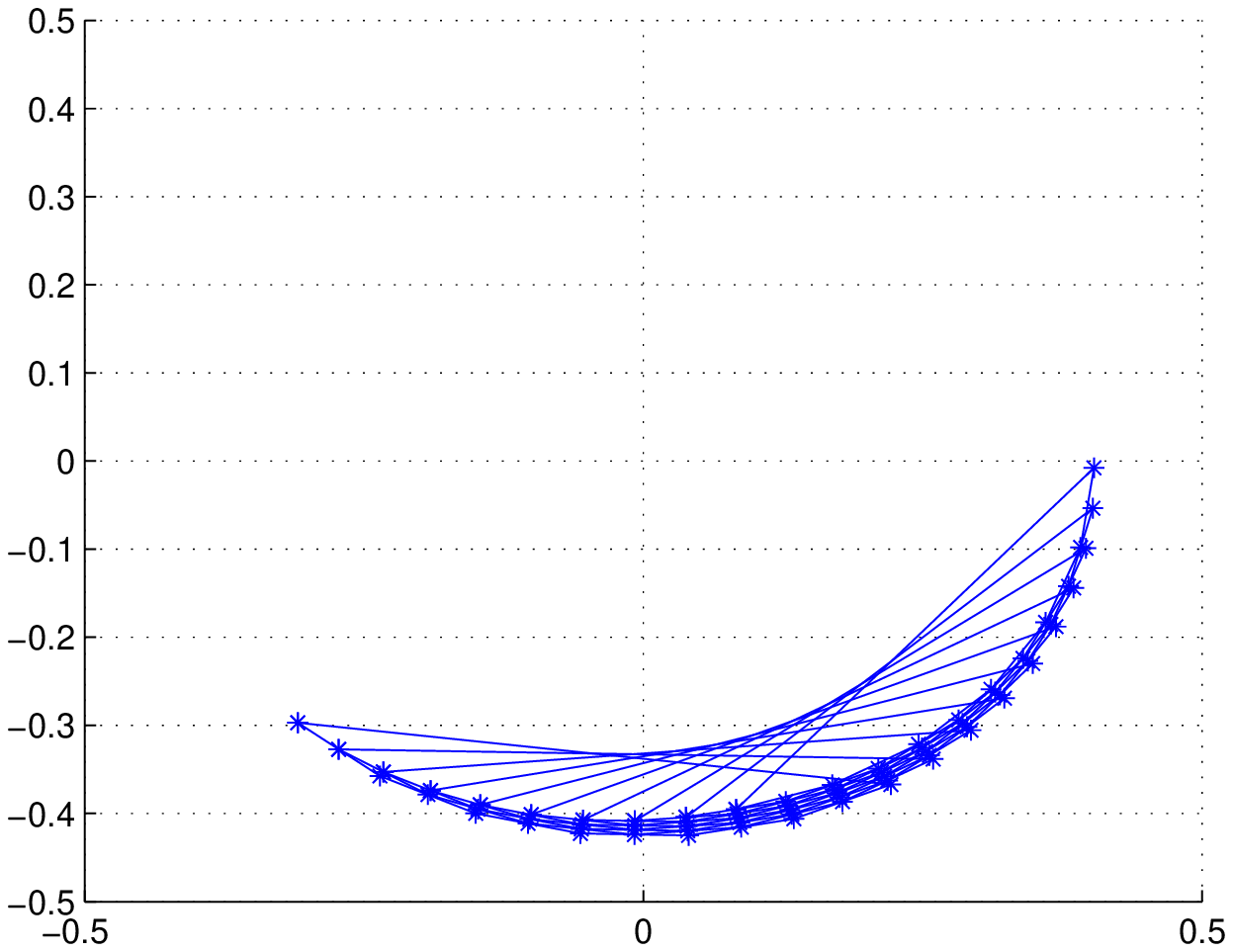}
\caption{$\lambda=-i$}
\label{fig:gamma_i}
\end{figure}

\begin{figure}
\hspace{-1cm}
\includegraphics[scale=0.4]{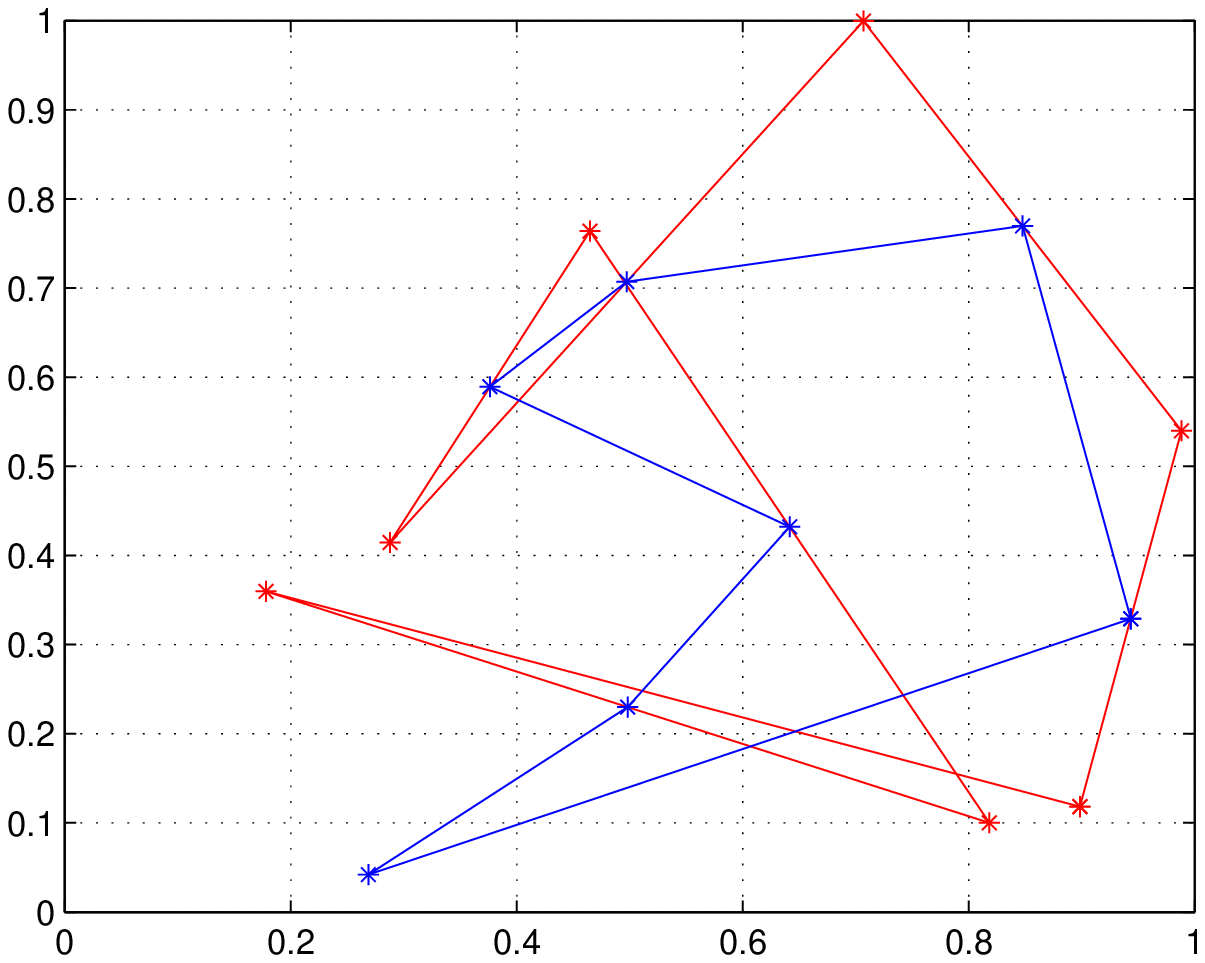}\includegraphics[scale=0.4]{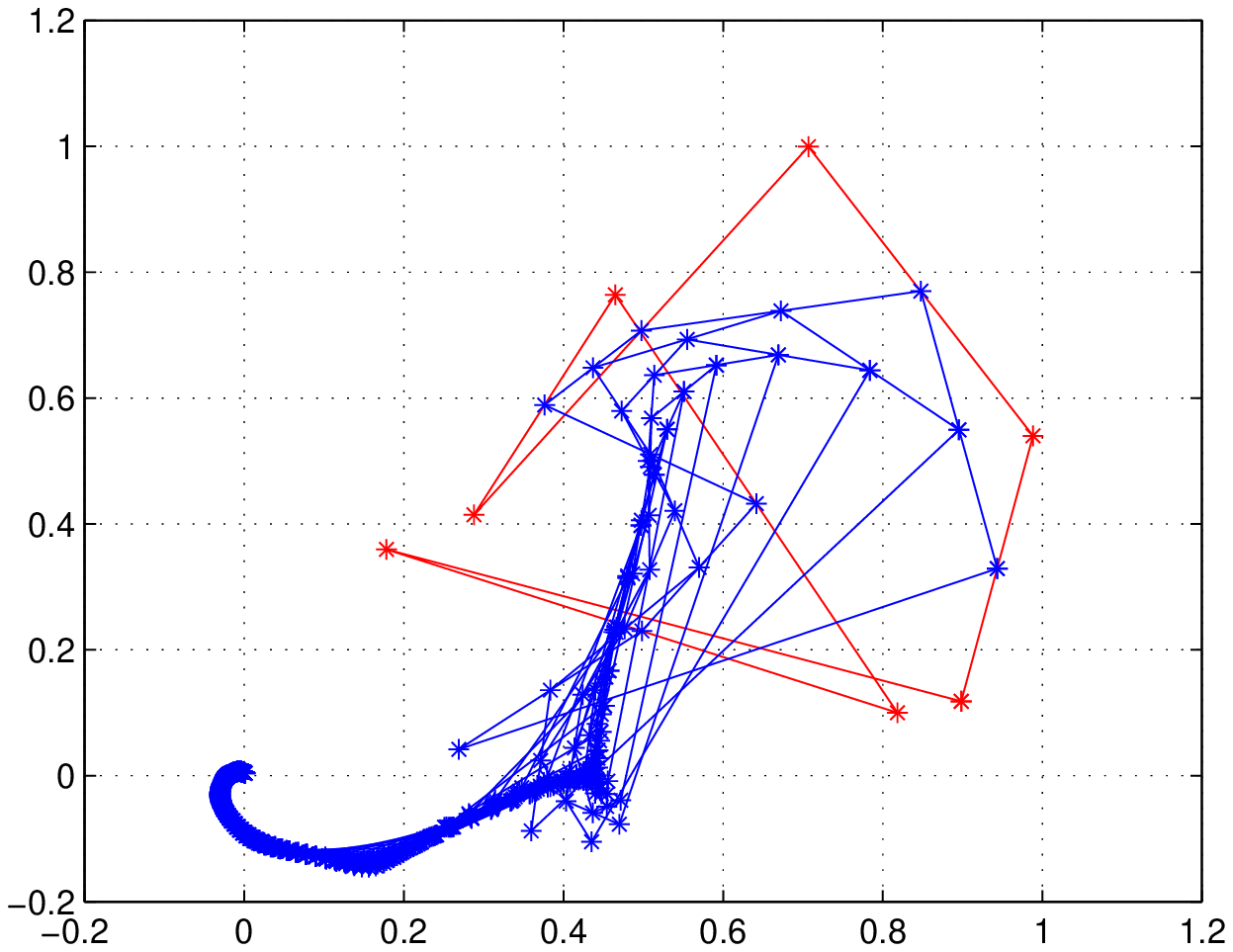}\includegraphics[scale=0.4]{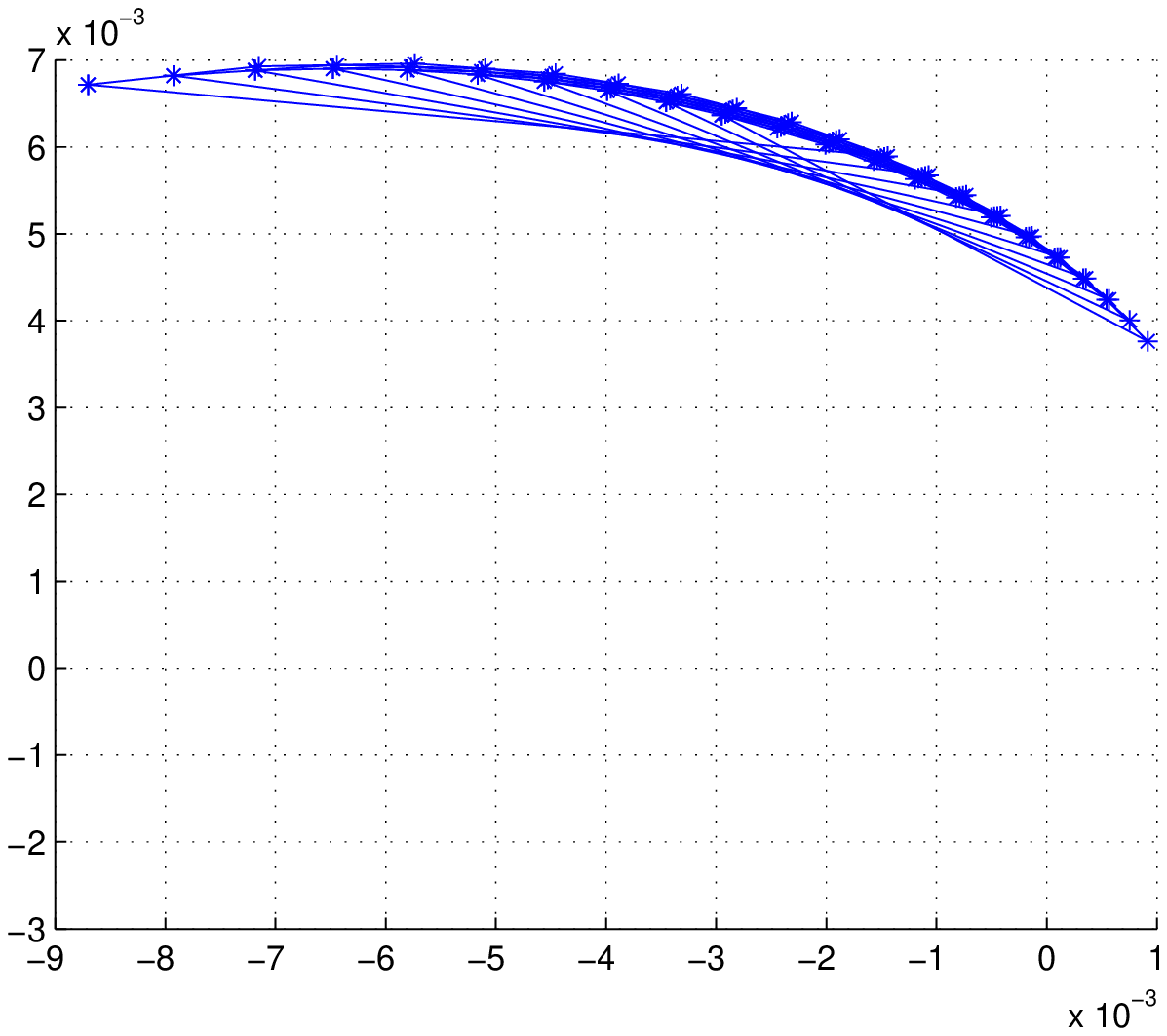}
\caption{$\lambda=\exp(i \pi/4)/2$}
\label{fig:gammacomp}
\end{figure}

\begin{figure}
\hspace{-1cm}
\includegraphics[scale=0.4]{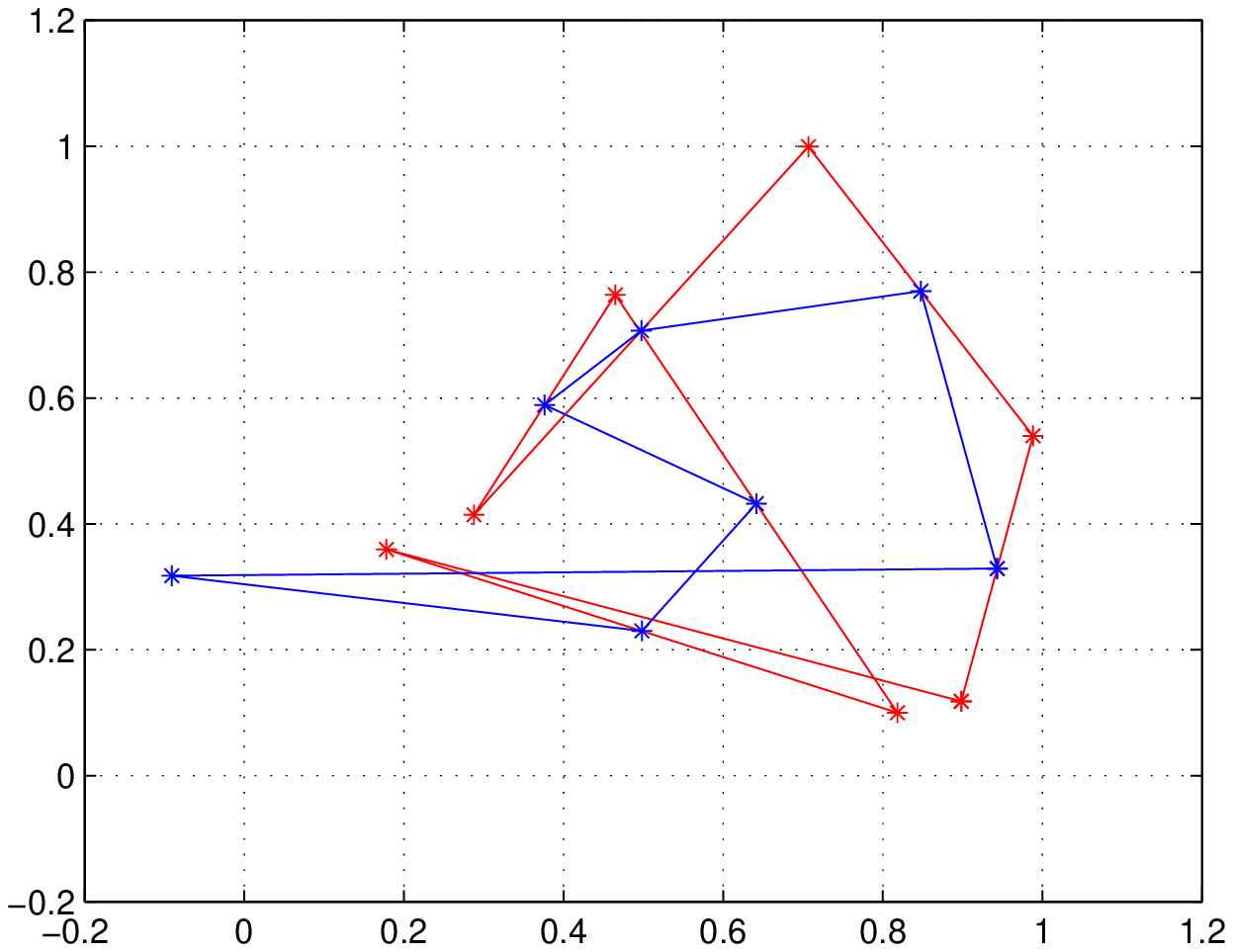}\includegraphics[scale=0.4]{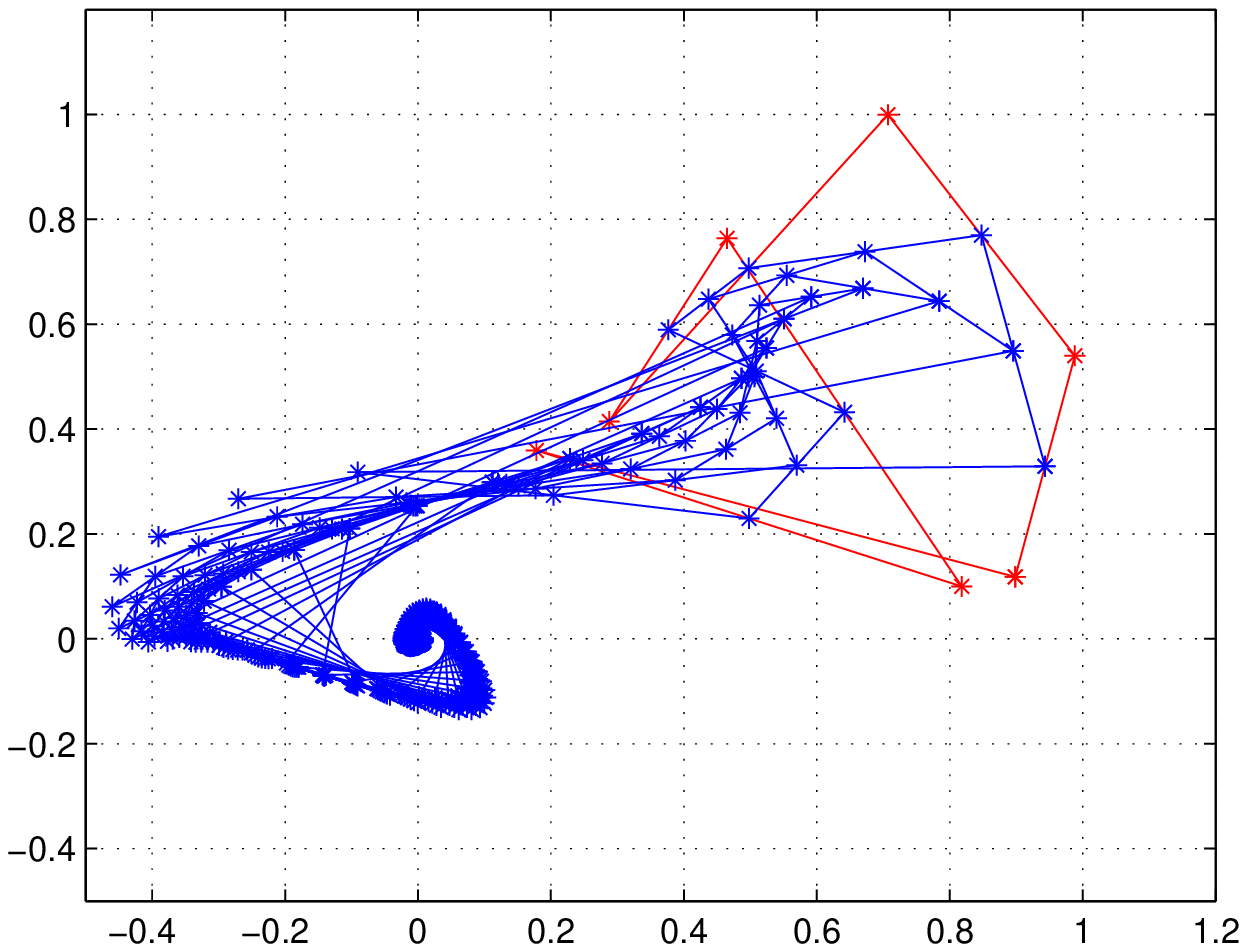}\includegraphics[scale=0.4]{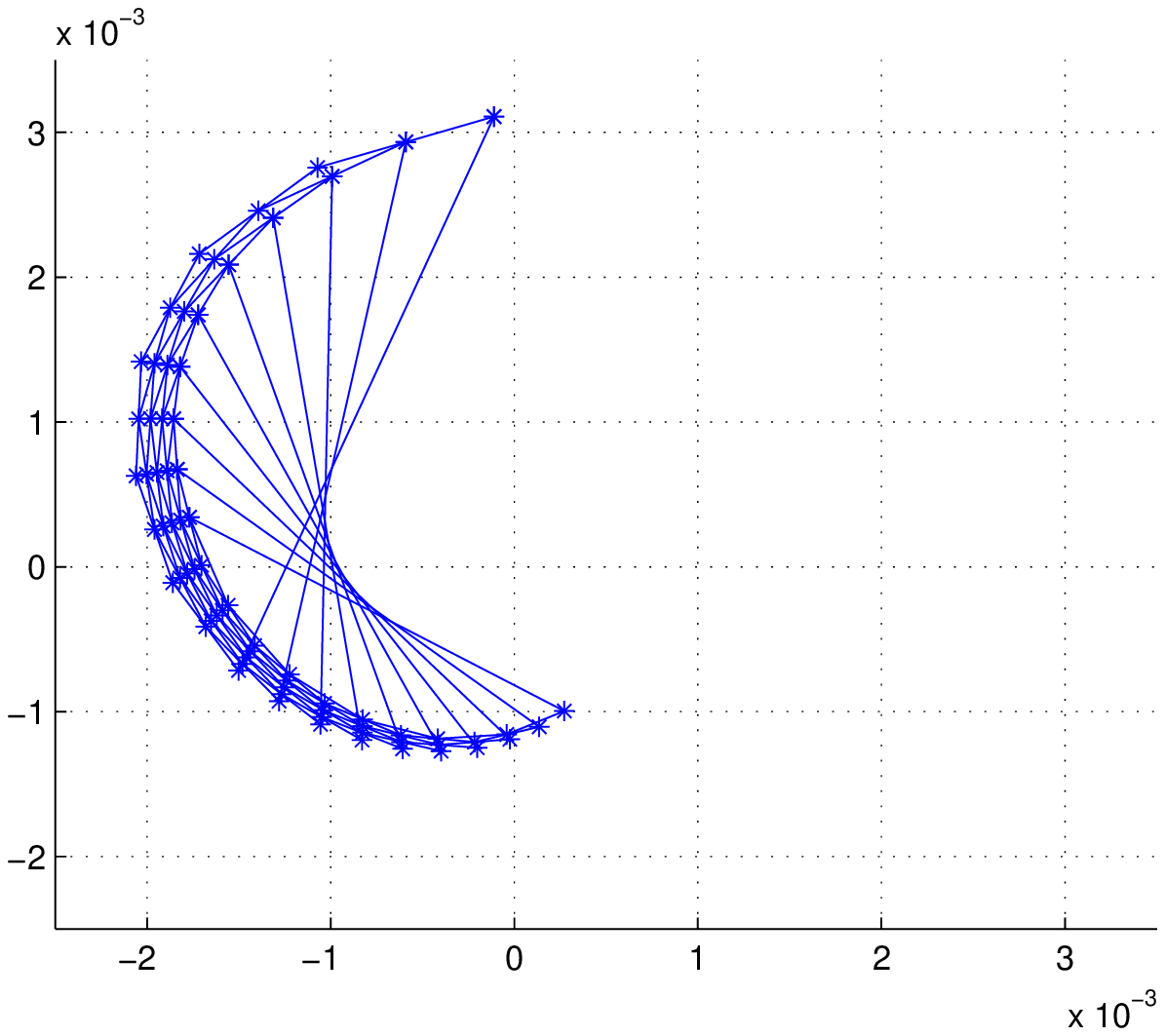}
\caption{$\lambda=\exp(i \pi/4 +i\pi)/2$}
\label{fig:gamma_comp}
\end{figure}

For the general case where $\lambda$ is some real or complex number, we 
have that
\begin{eqnarray*}
\Pm(t)_{t\rightarrow\infty}
&=&\!\!\!\!\!
\left[
\begin{array}{cccc}
1&&&\\
& \lambda^{1/N}&&\\
&&\ddots&\\
&&&\lambda^{\frac{N-1}{N}}
\end{array}
\right][\mathbf{FT}]
\left[
\begin{array}{cccc}
\mu_0^t&&&\\
&\mu_1^t&0&\\
&0&\ddots&\\
&&&\mu_{N-1}^t
\end{array}\right]_{t\rightarrow\infty}\!\!\!\!\!
\!\!\!\!\!\widetilde{\Pm}(0)\\
&=&\!\!\!\!\!
\left[
\begin{array}{cccc}
1&&&\\
& \lambda^{1/N}&&\\
&&\ddots&\\
&&&\lambda^{\frac{N-1}{N}}
\end{array}
\right][\mathbf{FT}]
\left[
\begin{array}{cccc}
\mu_0^t&&&\\
& 0&0&\\
&0&\ddots&\\
&&&0
\end{array}\right]_{t\rightarrow\infty}\!\!\!\!\!
\!\!\!\!\!\widetilde{\Pm}(0),
\end{eqnarray*}
where $\mu_0=\frac{1}{2}(1+\lambda^{1/N})$ is the dominant eigenvalue.
Since 
\[
\widetilde{\Pm}(0)=
[\mathbf{FT}]^*
\left[
\begin{array}{cccc}
1&&&\\
& \lambda^{-1/N}&&\\
&&\ddots&\\
&&&\lambda^{\frac{-(N-1)}{N}}
\end{array}
\right]
\Pm(0),
\]
we then have that
\begin{eqnarray*}
\Pm(t)_{t\rightarrow\infty}\!\!\!\!&=&\!\!\!\!
\mu_0^t
\left[
\begin{array}{cccc}
1&&&\\
& \lambda^{1/N}&&\\
&&\ddots&\\
&&&\lambda^{\frac{N-1}{N}}
\end{array}
\right]
[\mathbf{FT}]_{l,1}
([\mathbf{FT}]_{l,1})^*
\left[
\begin{array}{cccc}
1&&&\\
& \lambda^{-1/N}&&\\
&&\ddots&\\
&&&\lambda^{\frac{-(N-1)}{N}}
\end{array}
\right]
\Pm(0)
\end{eqnarray*}
and since the first column of the Fourier transform is 
a vector of all ones, this further simplifies to
\[
\Pm(t)_{t\rightarrow\infty}
=
\mu_0^t
\frac{1}{N}
\left[
\begin{array}{c}
1\\
\lambda^{1/N}\\
\vdots\\
\lambda^{\frac{N-1}{N}}
\end{array}
\right]
[
1,\lambda^{-1/N},\ldots,\lambda^{\frac{-(N-1)}{N}}
]
\Pm(0).
\]
Therefore, we see that the limiting behavior is dominated by
\[
\Pm(t)_{t\rightarrow\infty}
=
\left[
\frac{1}{2}
(1+\lambda^{1/N})
\right]^t
\frac{1}{N}
\underbrace{
[
1,\lambda^{-1/N},\ldots,\lambda^{\frac{-(N-1)}{N}}
]
\Pm(0)}_{\mbox{a (complex) scalar}}
\left[
\begin{array}{c}
1\\
\lambda^{1/N}\\
\vdots\\
\lambda^{\frac{N-1}{N}}
\end{array}
\right]
.
\]
We can distinguish different behaviors depending on $\lambda$.
\begin{enumerate}
\item
if $\lambda$ is real and $|\lambda|<1$, $\Pm(t)$ tends to zero, 
but the limit behavior will be a linear constellation of points
\[
(\alpha_t)_x
\left[
\begin{array}{c}
1\\
\lambda^{1/N}\\
\vdots\\
\lambda^{\frac{N-1}{N}}
\end{array}
\right]
+
i(\alpha_t)_y
\left[
\begin{array}{c}
1\\
\lambda^{1/N}\\
\vdots\\
\lambda^{\frac{N-1}{N}}
\end{array}
\right].
\]
If $|\lambda|>1$, the constellation of agent locations will diverge 
in a similar formation.
\item
If $\lambda$ is a complex number $\rho_{(\lambda)}e^{i\varphi_{(\lambda)}}$, 
the convergence/divergence will depend on the angle of rotation 
induced by $\varphi_{(\lambda)}$ and on the magnitude $\rho_{(\lambda)}$.
As seen in the examples provided in Figures 
\ref{fig:gamma01}, \ref{fig:gamma_1}, \ref{fig:gammai}, \ref{fig:gamma_i}, 
\ref{fig:gammacomp}, \ref{fig:gamma_comp}, in the limit, agents 
are marching in elliptic or circular arcs, spiralling towards their point 
of convergence (and in case of divergence, spiralling out to infinity).
As in Figure \ref{fig:gamma1}, the left figure presents 
the initial configuration (in red) and the first iteration (in blue), the 
second shows the entire evolution for 100 iterations (unless stated otherwise),
the last figure displays the scaled up configuration for the last 
few iterations.
\end{enumerate}

\subsection{Centroid gathering evolution and extensions}

As a second example, suppose that agent $\mathcal{P}_k$ is moving
according to the following linear combination of its own position, 
the positions of agents higher in the hierarchy i.e. $\{\mathcal{P}_{k+1},
\ldots, \mathcal{P}_{N-1}\}$ and the positions of those lower than
itself $\{\mathcal{P}_0, \mathcal{P}_1, \ldots, \mathcal{P}_{k-1}\}$:
\[
\mathcal{P}_k(t+1)= \alpha \mathcal{P}_k(t) +
\beta_F\sum_{l=k+1}^{N-1}\mathcal{P}_l(t)
+\beta_B\sum_{l=0}^{k-1}\mathcal{P}_l(t)
\]
or
\[
\Pm(t+1)= \left[
\begin{array}{ccccc}
\alpha&\beta_F&\beta_F&\ldots& \beta_F\\
\beta_B&\alpha&\beta_F&\ldots&\beta_F\\
&&\ddots&&\vdots\\
&&&&\\
\beta_B&\ldots&\ldots&\beta_B&\alpha
\end{array}\right]\Pm(t).
\]
Note that if $\beta_F=\beta_B=(1-\alpha)/(N-1)$, we will have
\begin{eqnarray*}
\mathcal{P}_k(t+1)
&=&\alpha \mathcal{P}_k(t) + \frac{1-\alpha}{N-1}\sum_{l=0,l\neq k}^N
   \mathcal{P}_l(t)\\
&=&\frac{N\alpha-1}{N-1}\mathcal{P}_k(t)
   +\left(1-\frac{N\alpha-1}{\alpha-1} \right)\Pc_{centroid}
\end{eqnarray*}
hence all agents move towards the time-invariant centroid on straight lines.

For general $\beta_F$ and $\beta_B$, the above matrix is 
$\beta_B/\beta_F$-factor circulant and is diagonalized by 
\[
\mathbf{\widetilde{P}}(t)= [\mathbf{FT}]^* \left[
\begin{array}{cccc}
1&&&\\
&(\beta_B/\beta_F)&0&\\
&&\ddots&\\
&0&&(\beta_B/\beta_F)^{N-1}
\end{array}
\right]\Pm(t),
\]
the modes or eigenvalues being given by
\[
\mu_l=\alpha+\sum_{k=1}^{N-1}
\beta_F\left(\frac{\beta_B}{\beta_F}\right)^{\frac{k}{N}}e^{-i\frac{2\pi}{N}kl},
~l=0,\ldots,N-1.
\]

\begin{figure}
\hspace{-1cm}
\includegraphics[scale=0.4]{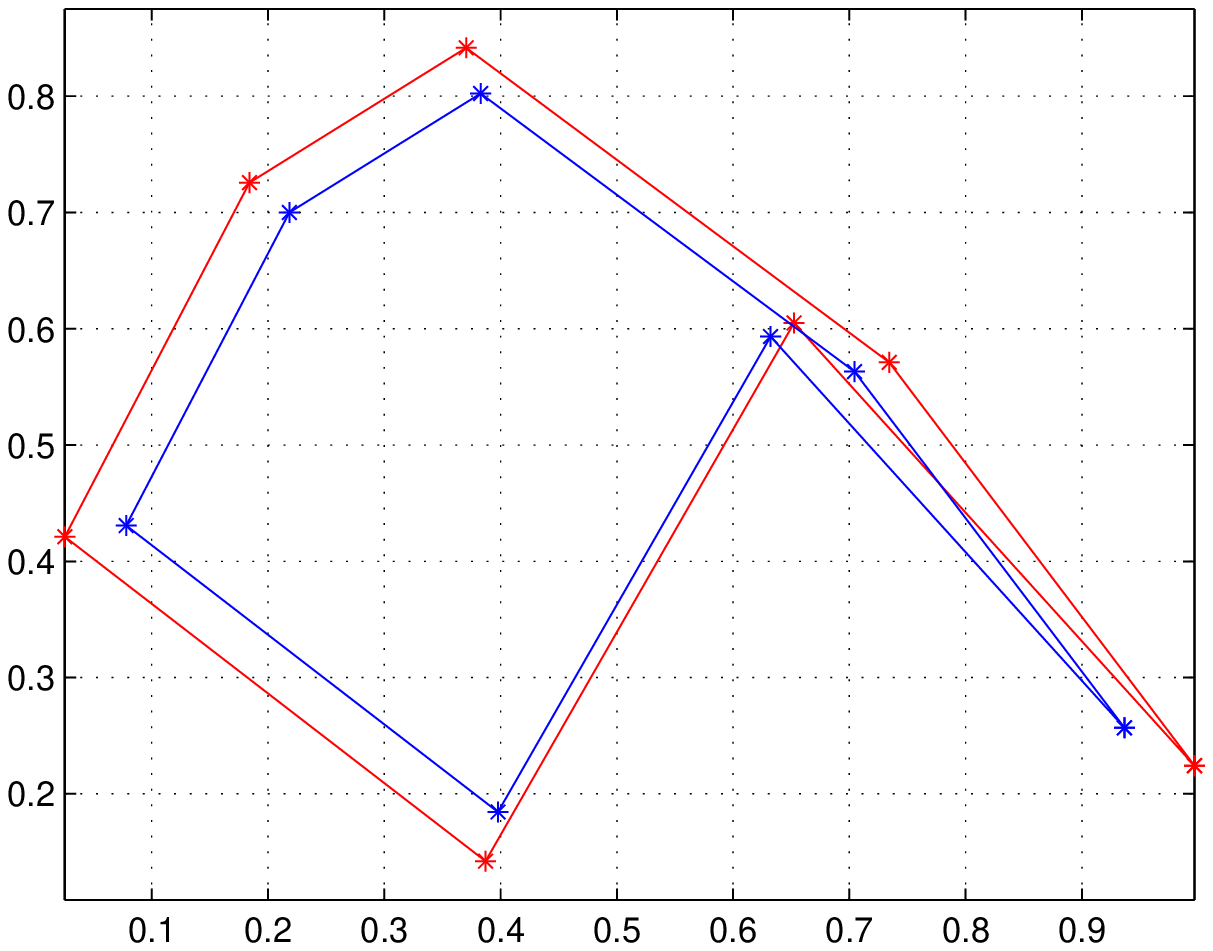}\includegraphics[scale=0.4]{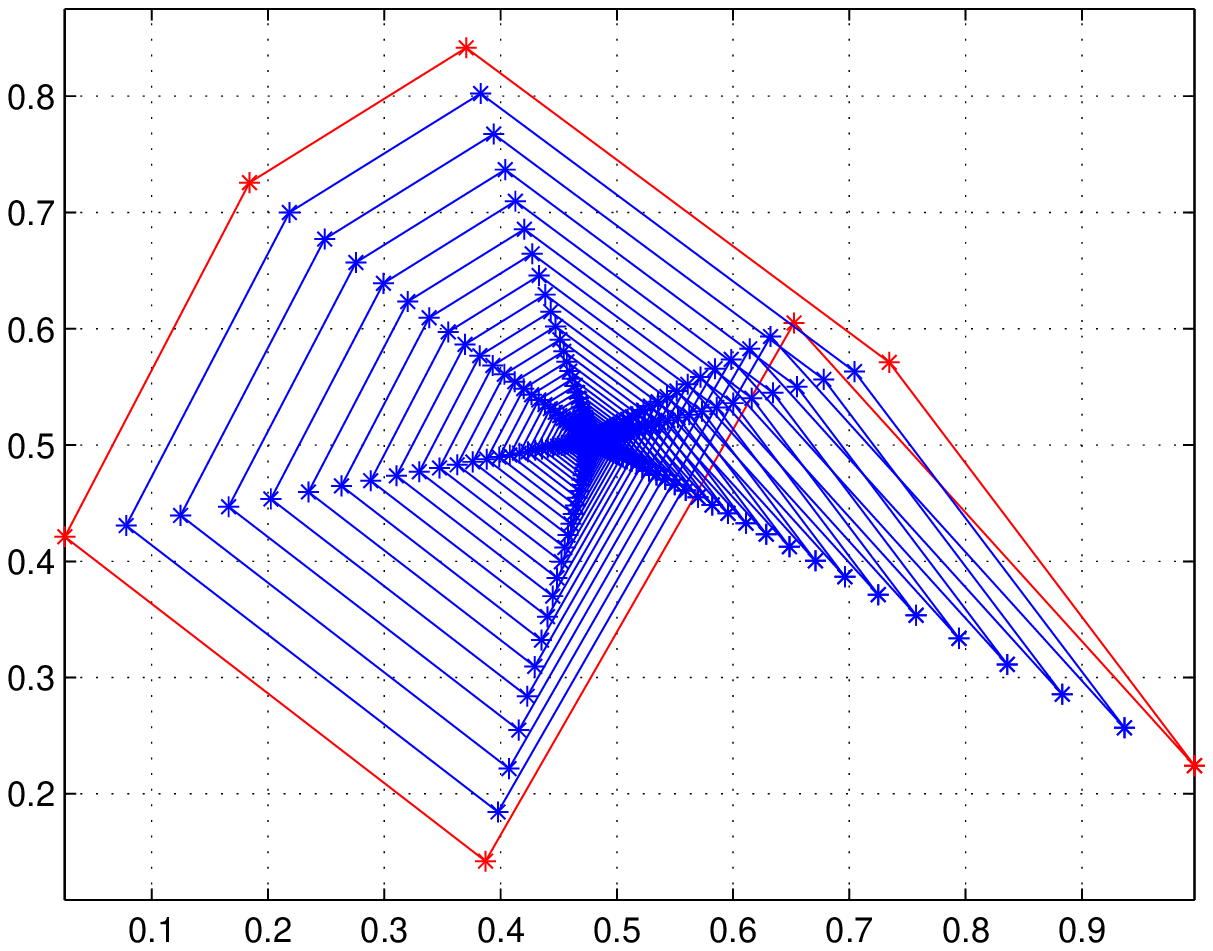}\includegraphics[scale=0.4]{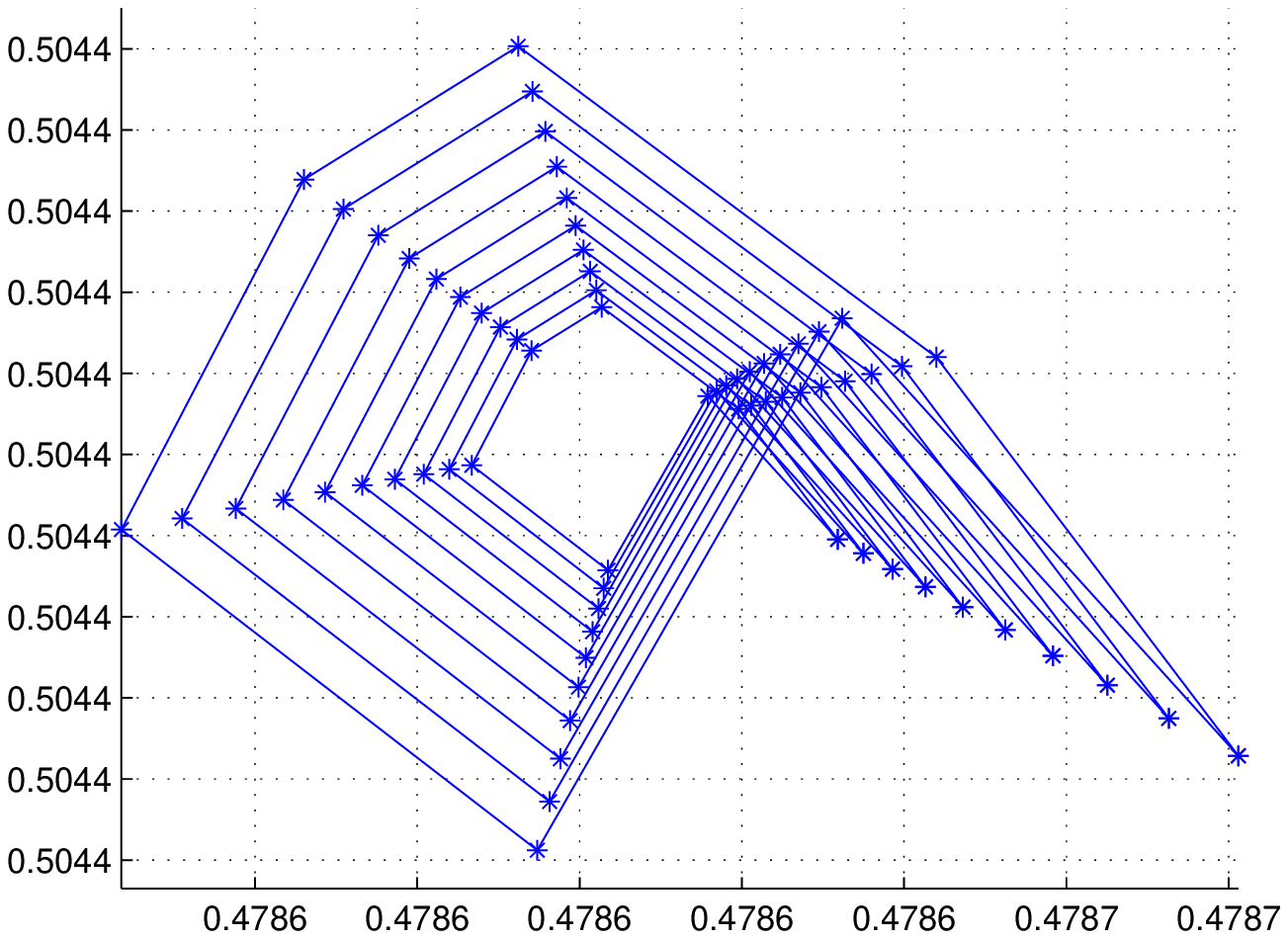}
\caption{$\lambda=1$, $\alpha=0.1$, 100 iterations}
\label{second:lambda1}
\end{figure}

Let us consider first the case of perfectly cyclic interaction, i.e., when 
$\beta_B=\beta_F$. In this case, the interaction matrix is circulant, 
and we have 
\[
\mu_l=\alpha+\sum_{k=1}^{N-1}\beta_Fe^{-i\frac{2\pi}{N}kl},
~l=0,\ldots,N-1
\]
and
\begin{eqnarray*}
\mu_0 &=& \alpha + (N-1)\beta_F \\
\mu_l &=& \alpha -\beta_F +\sum_{k=0}^{N-1}\beta_F e^{-i\frac{2\pi}{N}kl} 
= \alpha-\beta_F.
\end{eqnarray*}
For normalization, we shall take $\beta_F=(1-\alpha)/(N-1)$ and then
\begin{eqnarray*}
\mu_0 &=& 1 \\
\mu_l &=& (N\alpha -1)/(N-1),\mbox{ for all }l.
\end{eqnarray*}
We now have that 
\[
\widetilde{\Pm}(t)=[\mathbf{FT}]^*\Pm(t)
\]
evolves according to
\[
\widetilde{\Pm}(t)_{t\rightarrow\infty}
=
\left[
\begin{array}{cccc}
1&&&\\
& \left(\frac{N\alpha-1}{N-1}\right)^t&&\\
&&\ddots&\\
&&& \left(\frac{N\alpha-1}{N-1}\right)^t
\end{array}
\right]
\widetilde{\Pm}(0)\\
=
\left[
\begin{array}{c}
1 \\
0\\
\vdots \\
0
\end{array}
\right]
[1,0,\ldots,0]\widetilde{\Pm}(0).
\]
Hence
\[
\Pm(t)_{t\rightarrow\infty}
=
[\mathbf{FT}]
\left[
\begin{array}{c}
1 \\
0\\
\vdots \\
0
\end{array}
\right]
[1,0,\ldots,0]
[\mathbf{FT}]^*
\Pm(0)\\
=
\frac{1}{N}
\left[
\begin{array}{c}
1 \\
1\\
\vdots \\
1
\end{array}
\right]
[1,1,\ldots,1]
\Pm(0)
\]
i.e., as we have already seen, all points converge towards the centroid 
of the initial constellation. The convergence will be as follows:
\[
\Pm^N(t)_{t\rightarrow\infty}
=
\widetilde{\Pm}(t)-
\left[
\begin{array}{c}
1\\
0\\
\vdots\\
0
\end{array}
\right]
[
1,0,\ldots,0
]\widetilde{\Pm}(0)\\
=
\left(
\frac{N\alpha-1}{N-1}
\right)^t
\left[
\begin{array}{cccc}
0 &&&\\
 & 1 &&\\
&&\ddots &\\
0 &&& 1
\end{array}
\right]
\widetilde{\Pm}(0).
\]
Therefore
\begin{eqnarray*}
\Pm^N(t)_{t\rightarrow\infty}
&=&
[\mathbf{FT}]
\left(
I-
\left[
\begin{array}{c}
1\\
0\\
\vdots\\
0
\end{array}
\right]
[
1,0,\ldots,0
]
\right)
[\mathbf{FT}]^*
\widetilde{\Pm}(0)\\
&=&
\left(
\frac{N\alpha-1}{N-1}
\right)^t
\left(
\Pm(0)-
\left[
\begin{array}{c}
1\\
1\\
\vdots\\
1
\end{array}
\right]
[
1,1,\ldots,1
]
\Pm(0)
\right)
.
\end{eqnarray*}
Consequently, all agents will gather towards the centroid by moving 
on a line from $\Pc_k(0)$ to $(1/N)\sum_{i=1}^{N-1}\Pc_i(0)$ 
(see Figure \ref{second:lambda1}).

\begin{figure}
\hspace{-1cm}
\includegraphics[scale=0.4]{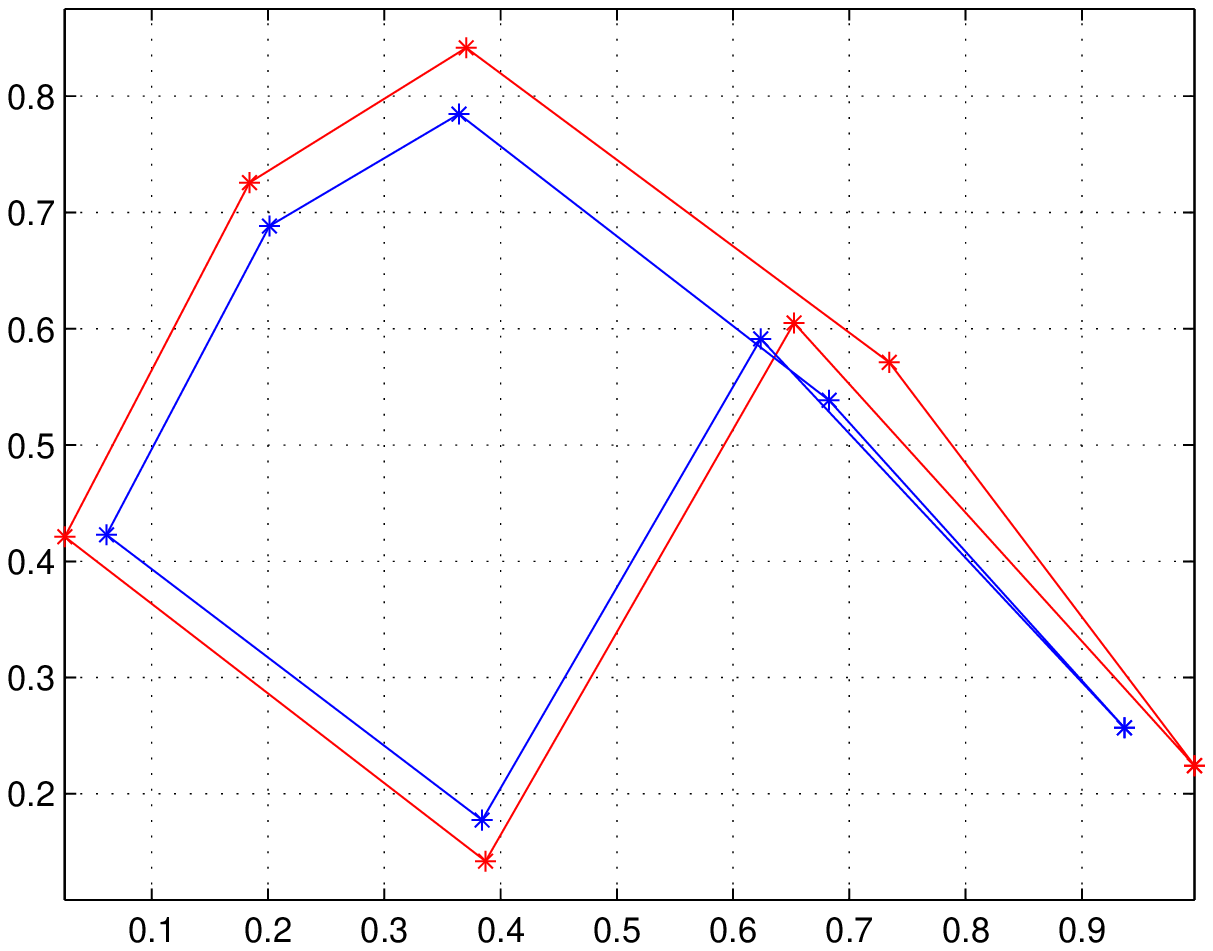}\includegraphics[scale=0.4]{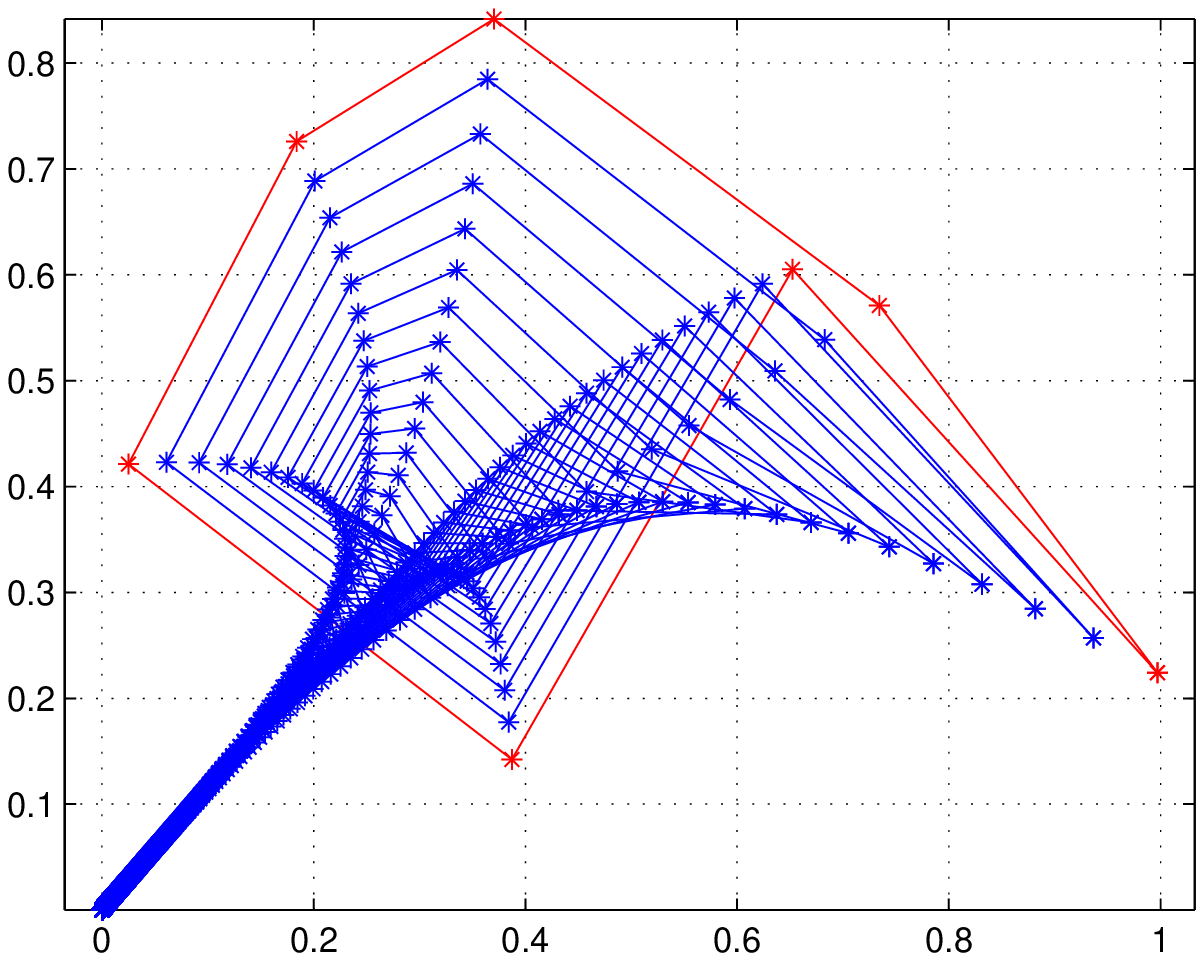}\includegraphics[scale=0.4]{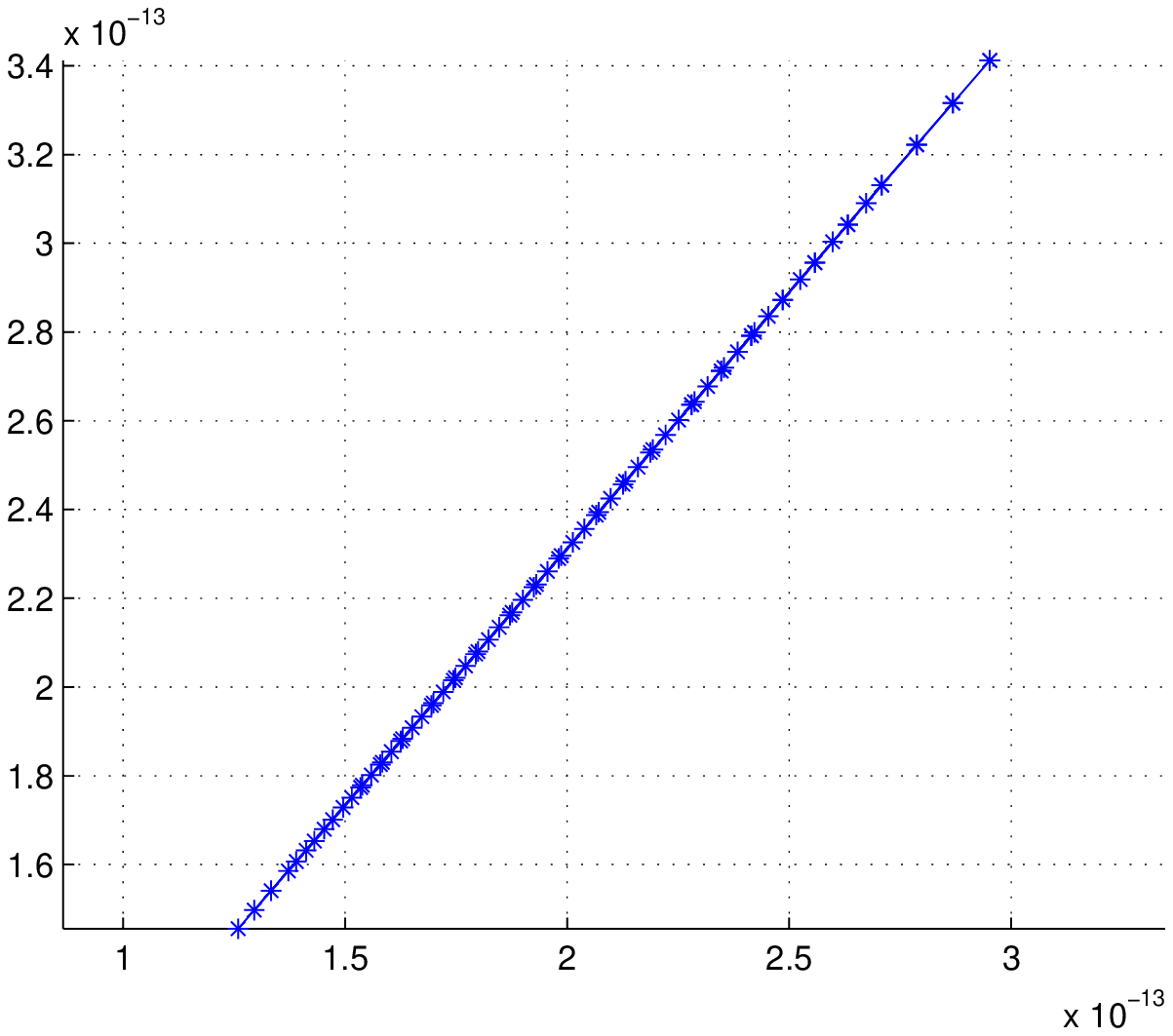}
\caption{$\lambda=0.5$, $\alpha=0.1$, 1000 iterations}
\label{fig:lambda01}
\end{figure}

\begin{figure}
\hspace{-1cm}
\includegraphics[scale=0.4]{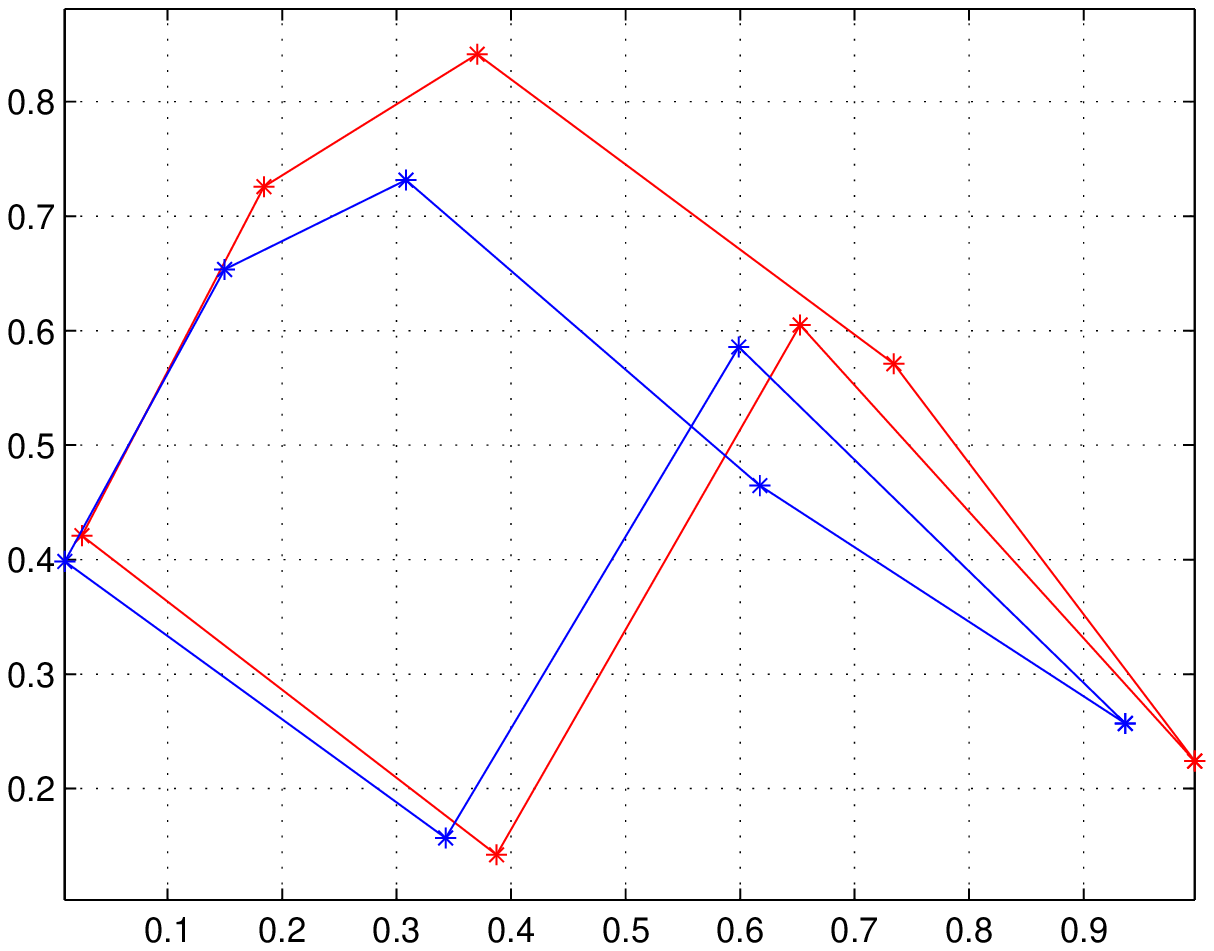}\includegraphics[scale=0.4]{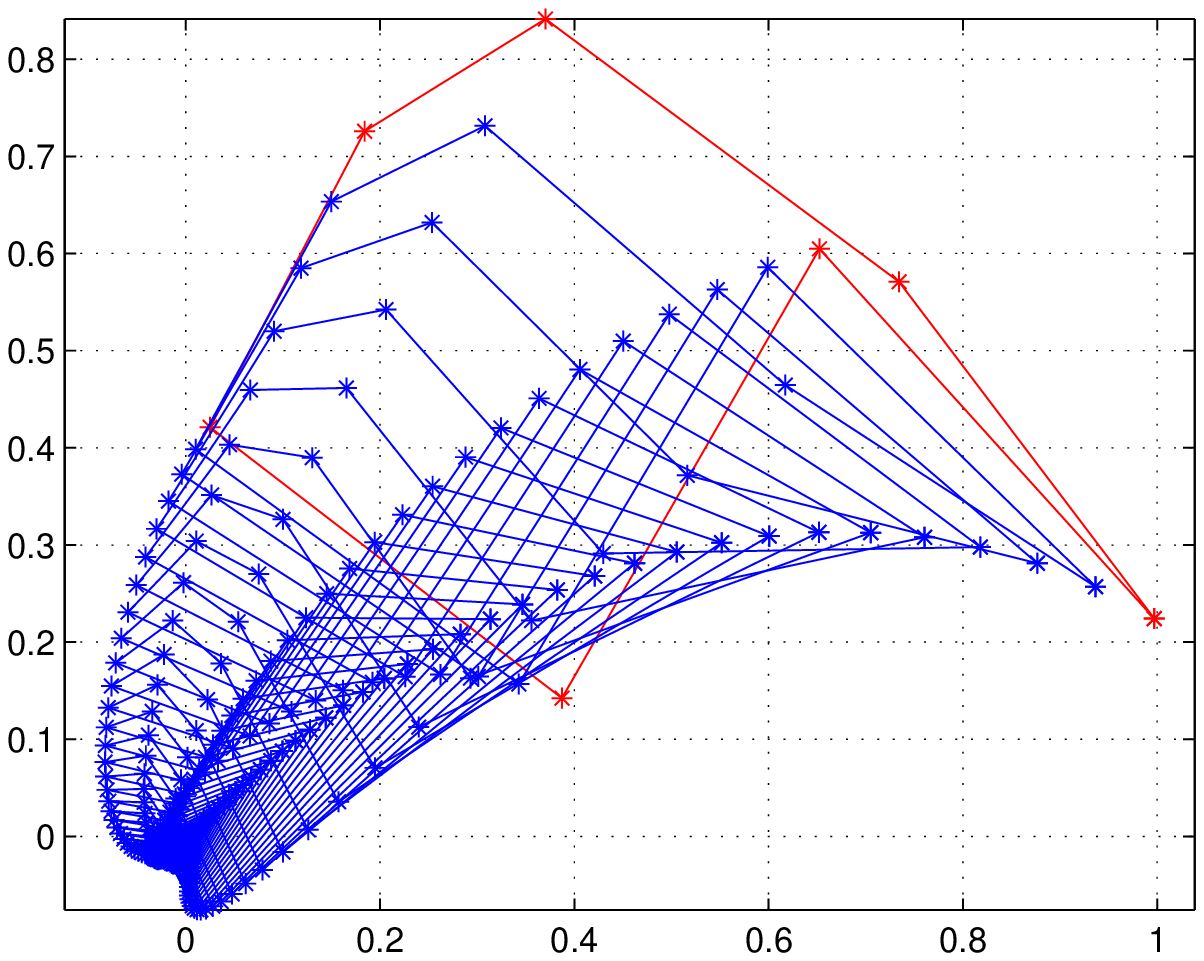}\includegraphics[scale=0.4]{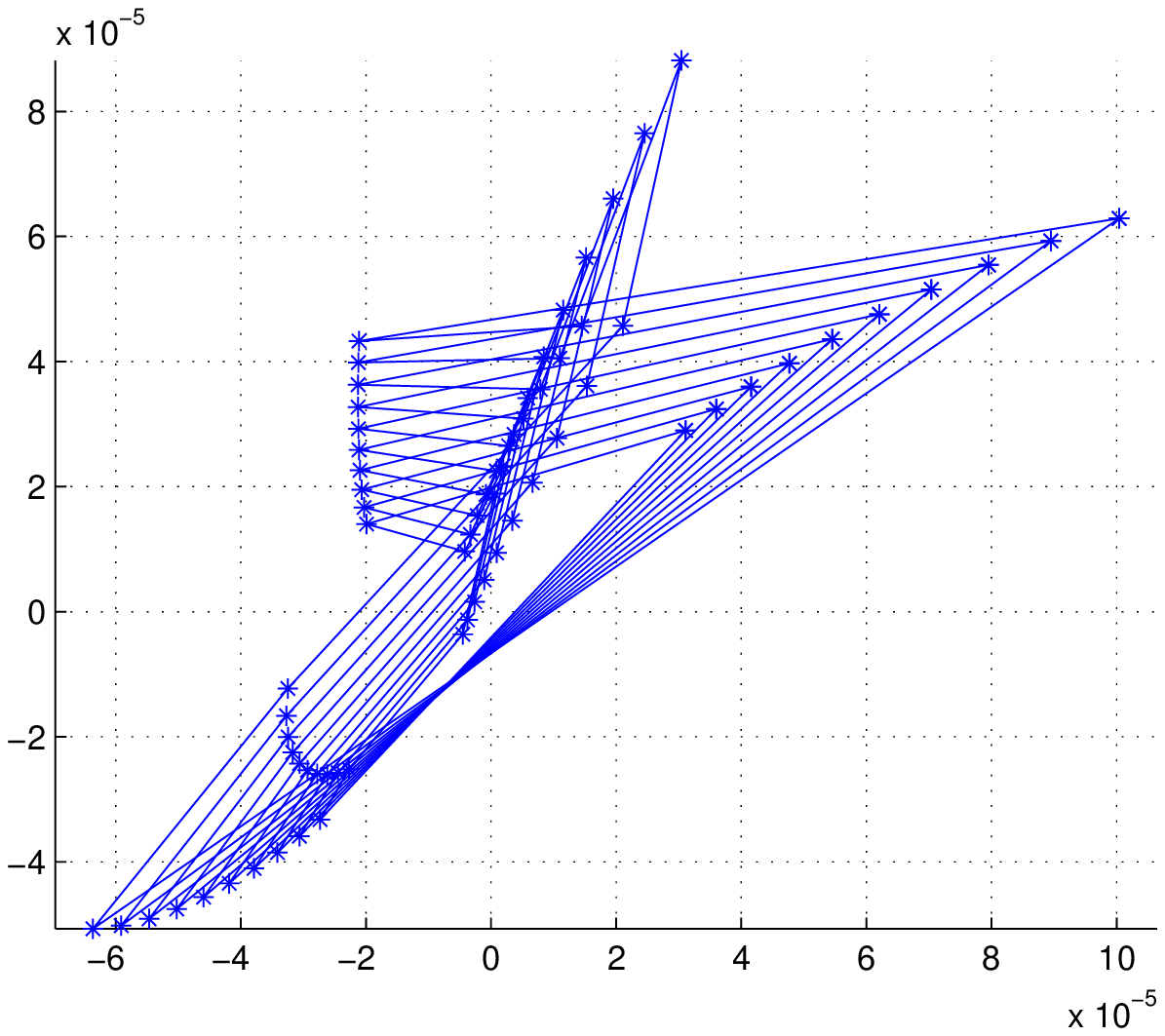}
\caption{$\lambda=-1$, $\alpha=0.1$, 100 iterations}
\label{fig:lambda_1}
\end{figure}

\begin{figure}
\hspace{-1cm}
\includegraphics[scale=0.4]{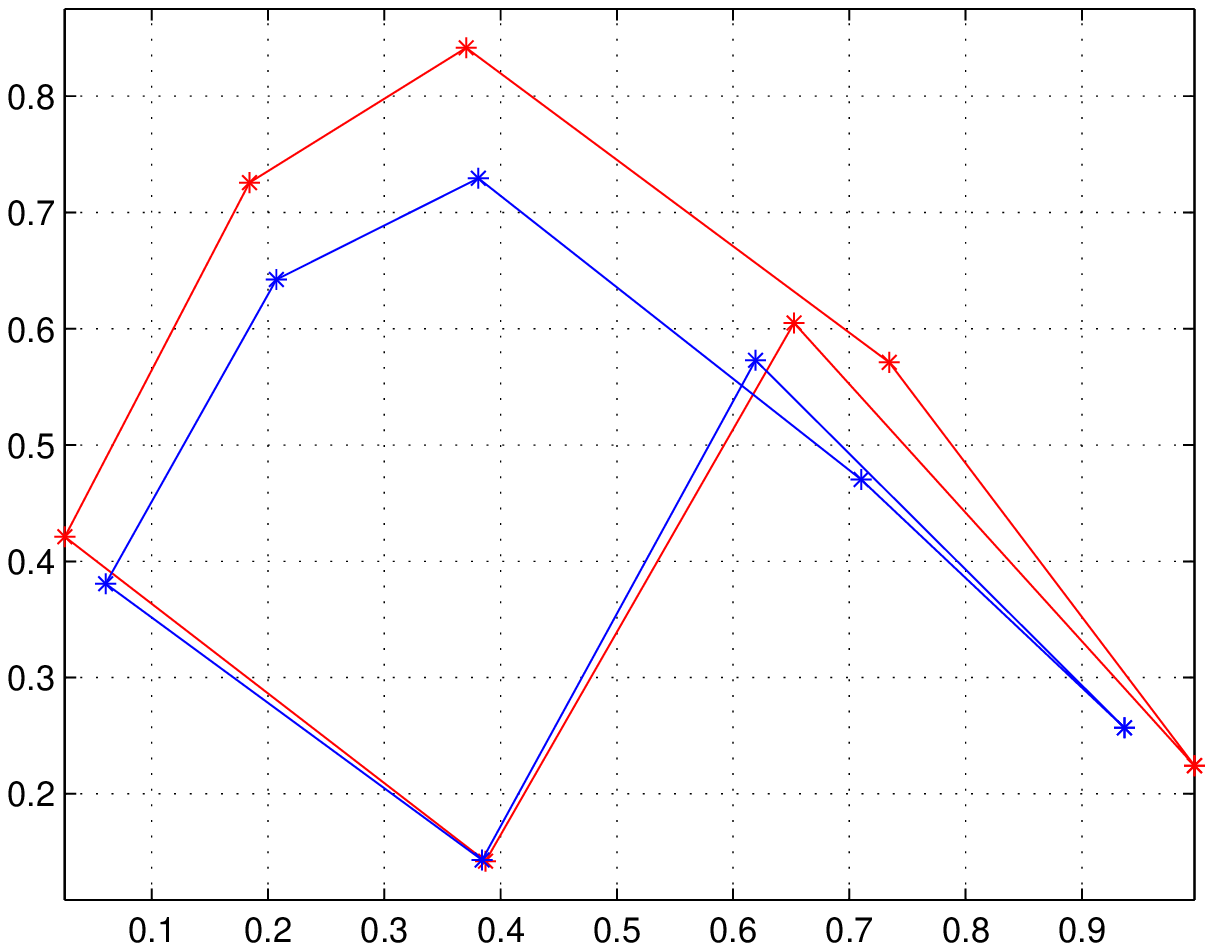}\includegraphics[scale=0.4]{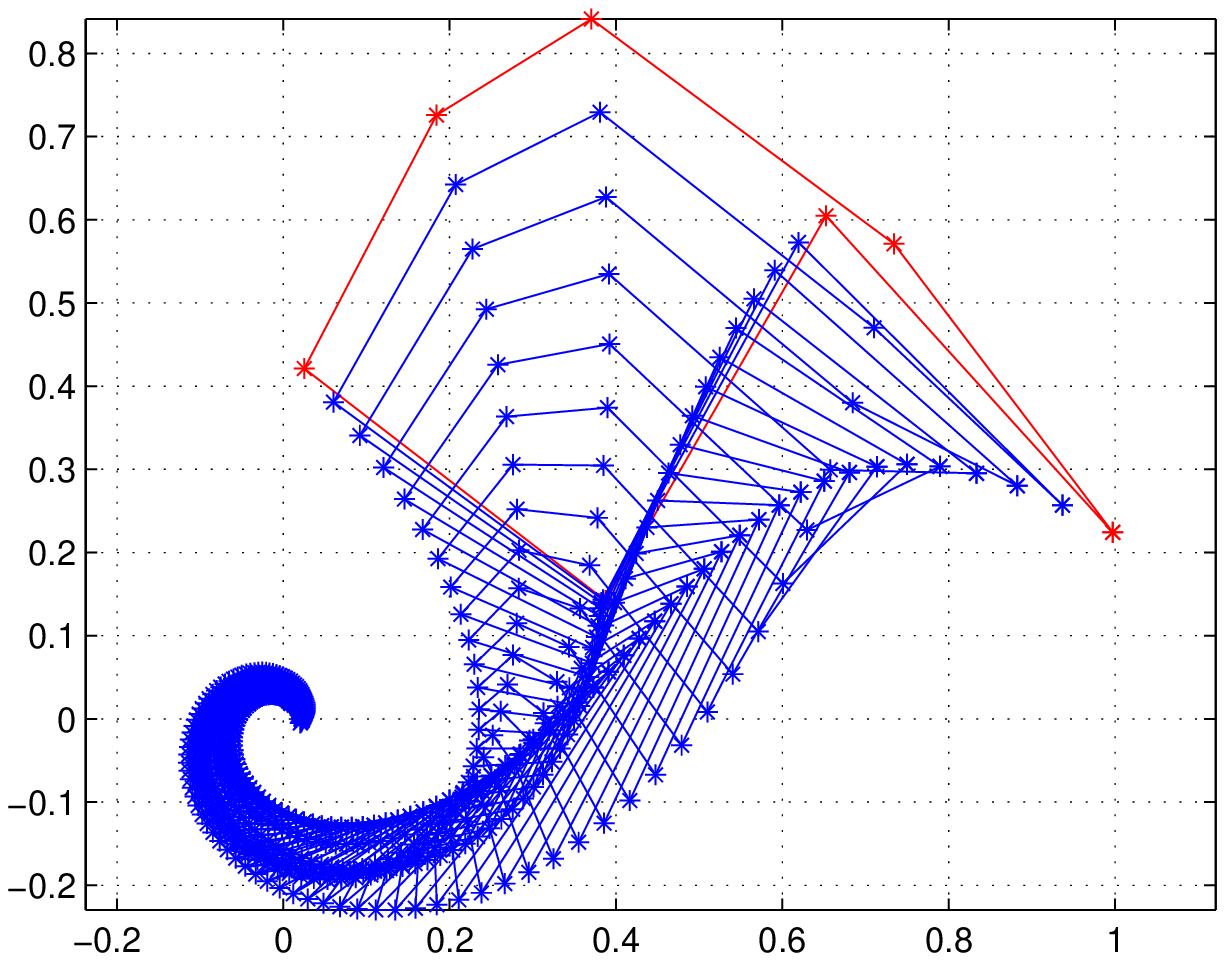}\includegraphics[scale=0.4]{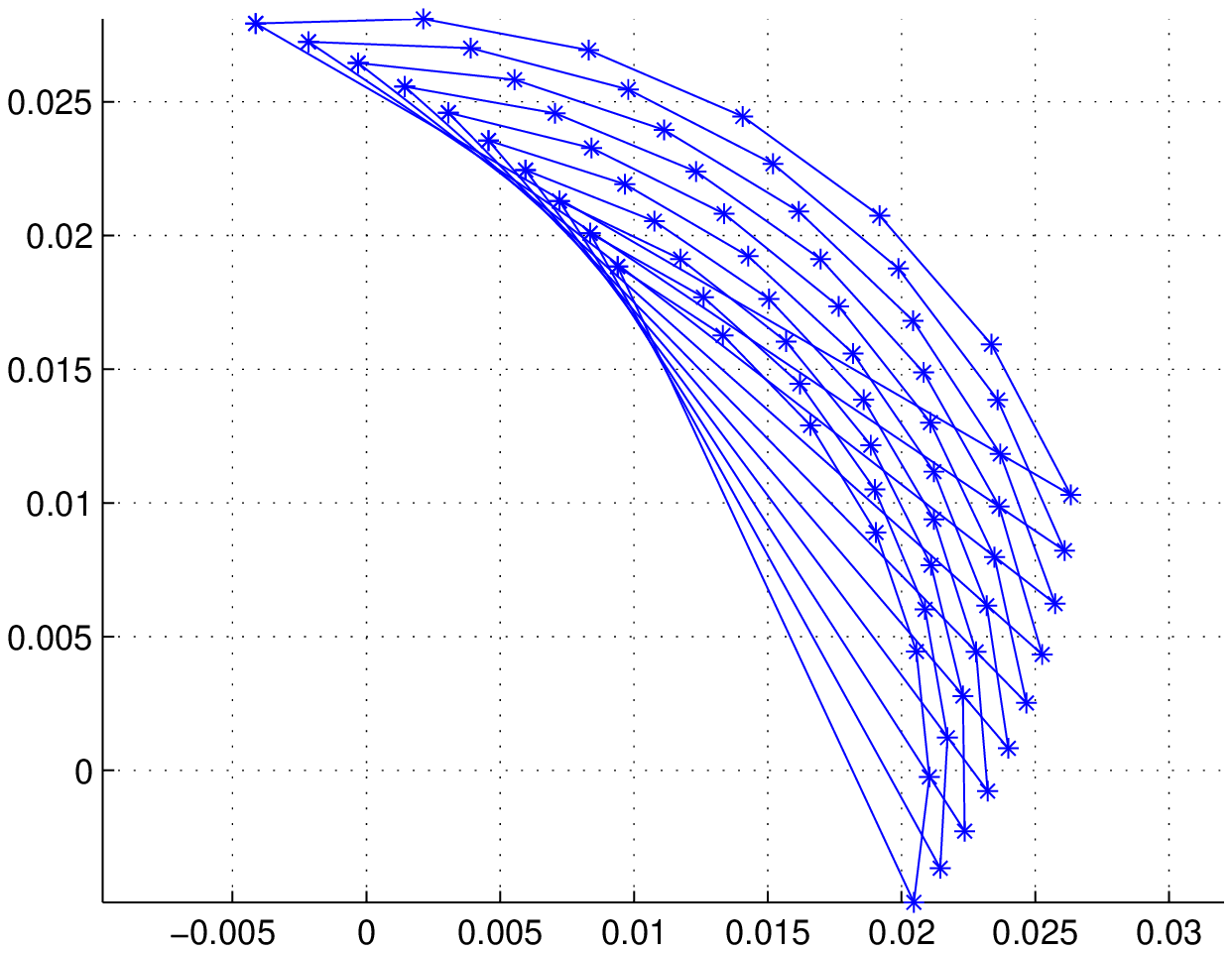}
\caption{$\lambda=i$, $\alpha=0.1$, 100 iterations}
\label{fig:lambdai}
\end{figure}

\begin{figure}
\hspace{-1cm}
\includegraphics[scale=0.4]{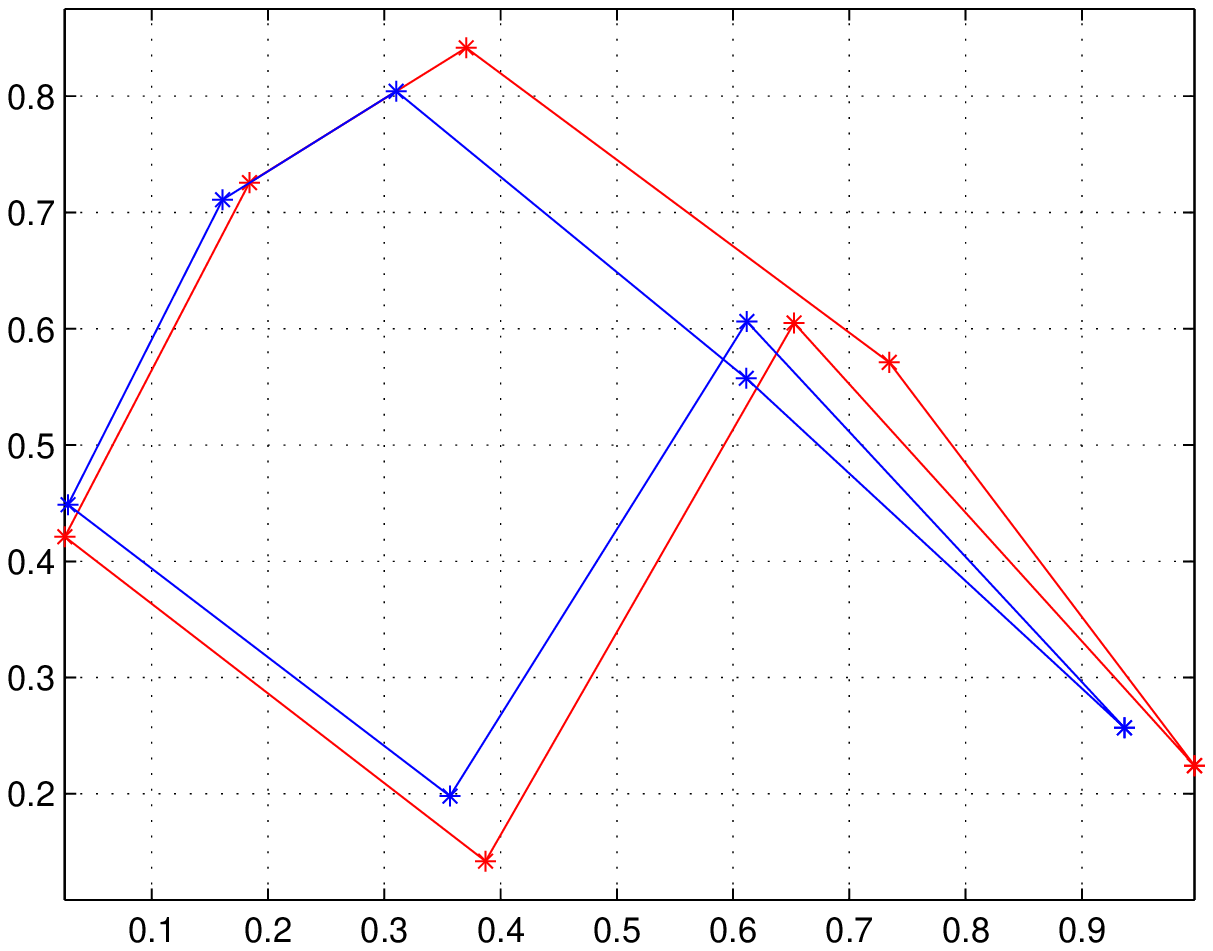}\includegraphics[scale=0.4]{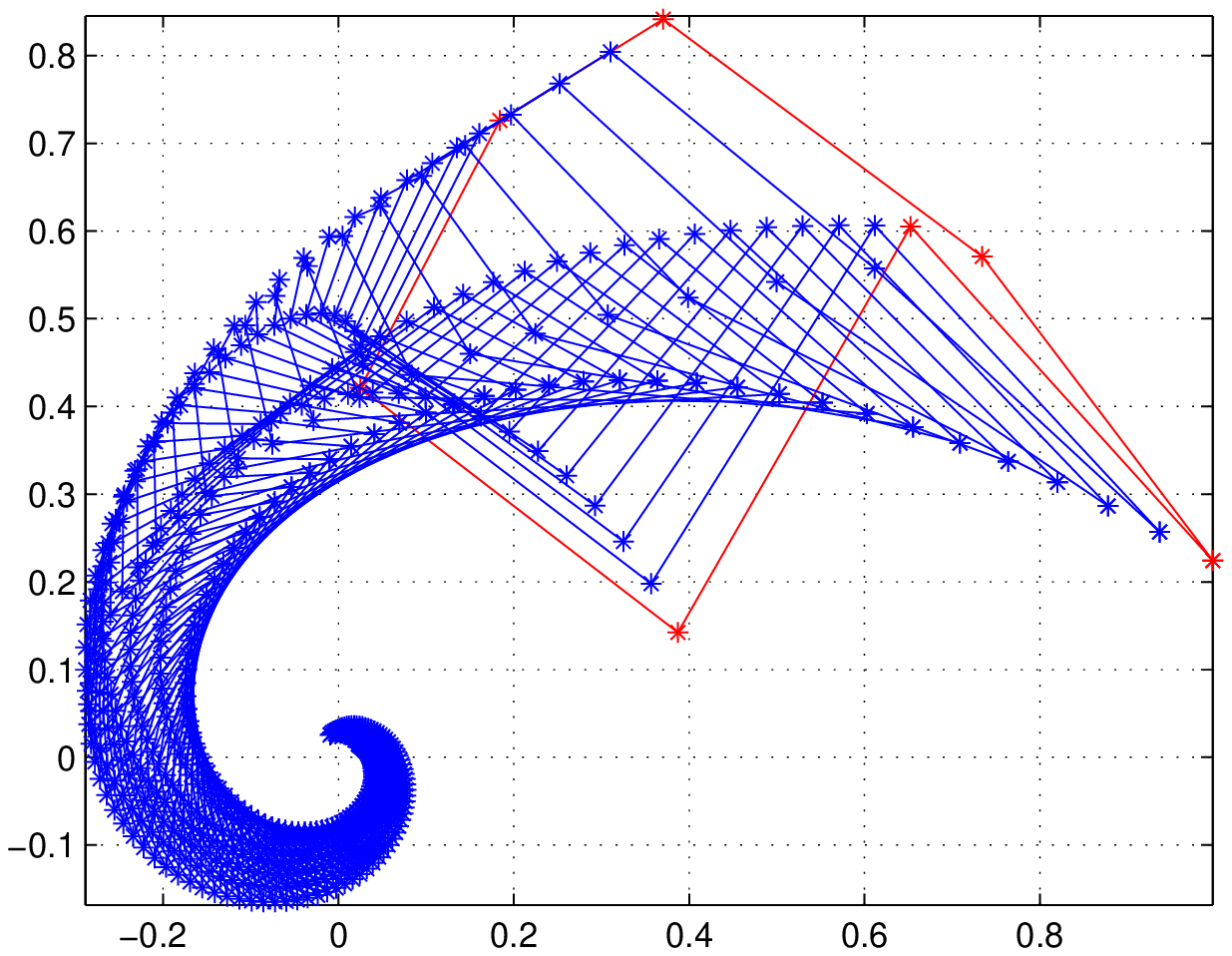}\includegraphics[scale=0.4]{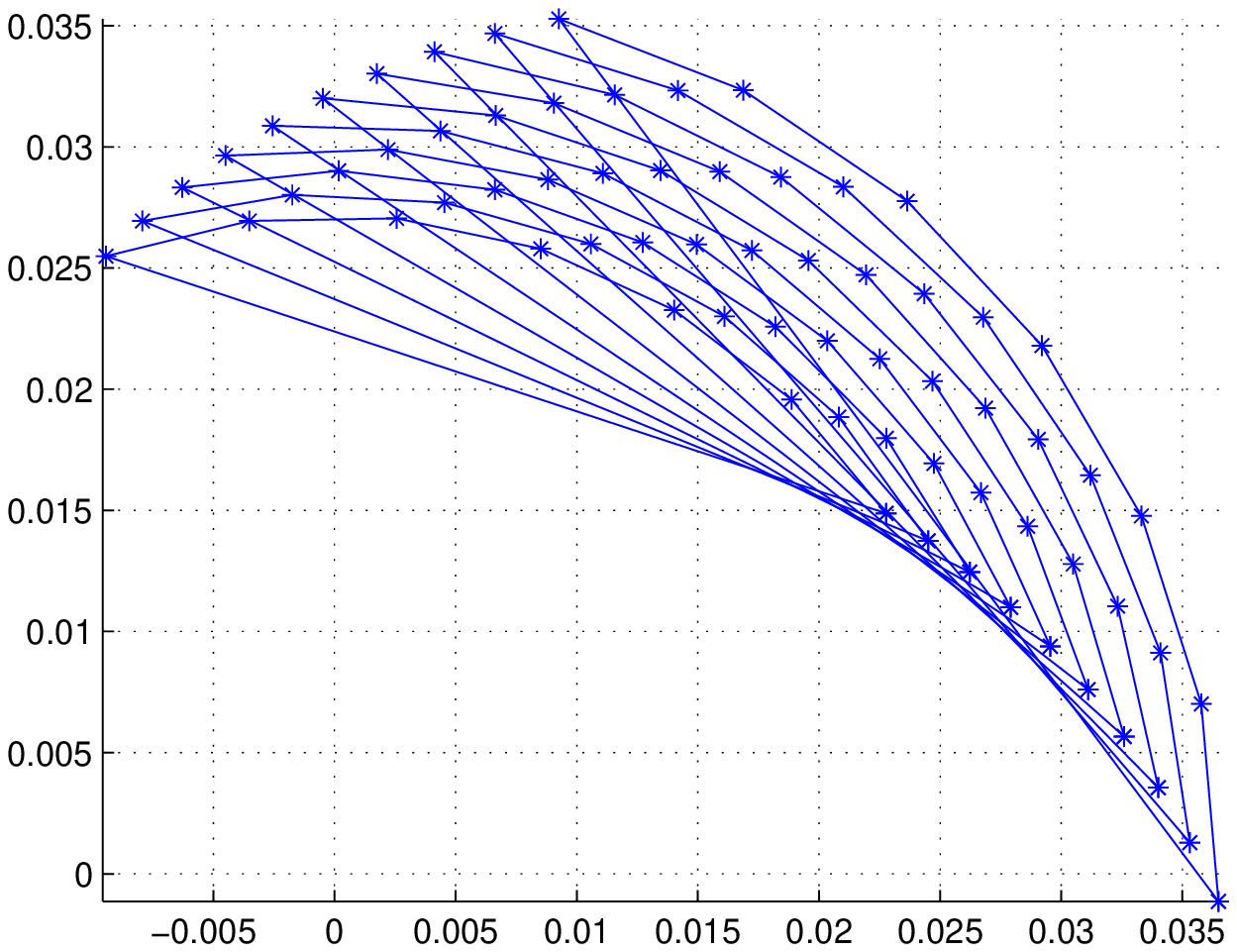}
\caption{$\lambda=-i$, $\alpha=0.1$, 100 iterations}
\label{fig:lambda_i}
\end{figure}

\begin{figure}
\hspace{-1cm}
\includegraphics[scale=0.4]{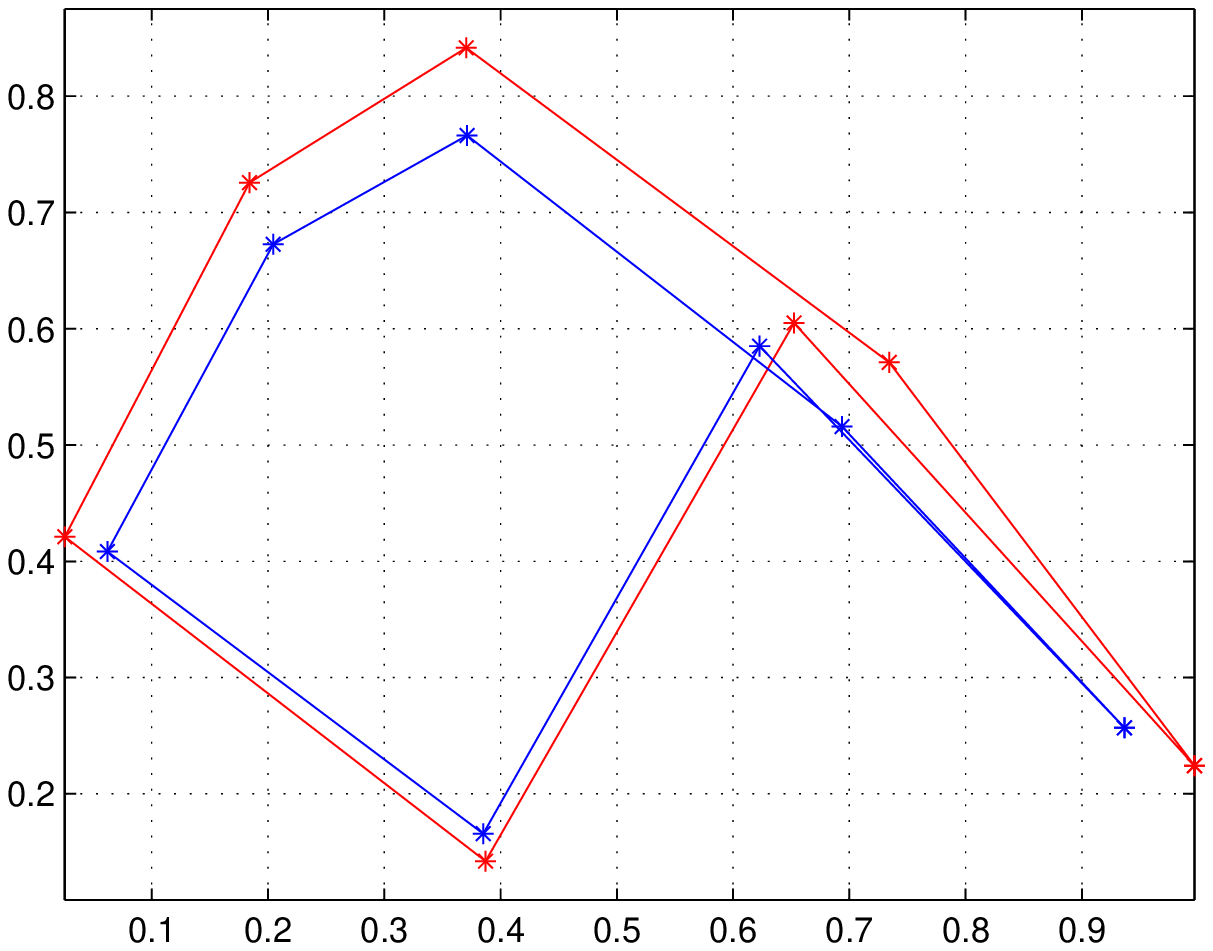}\includegraphics[scale=0.4]{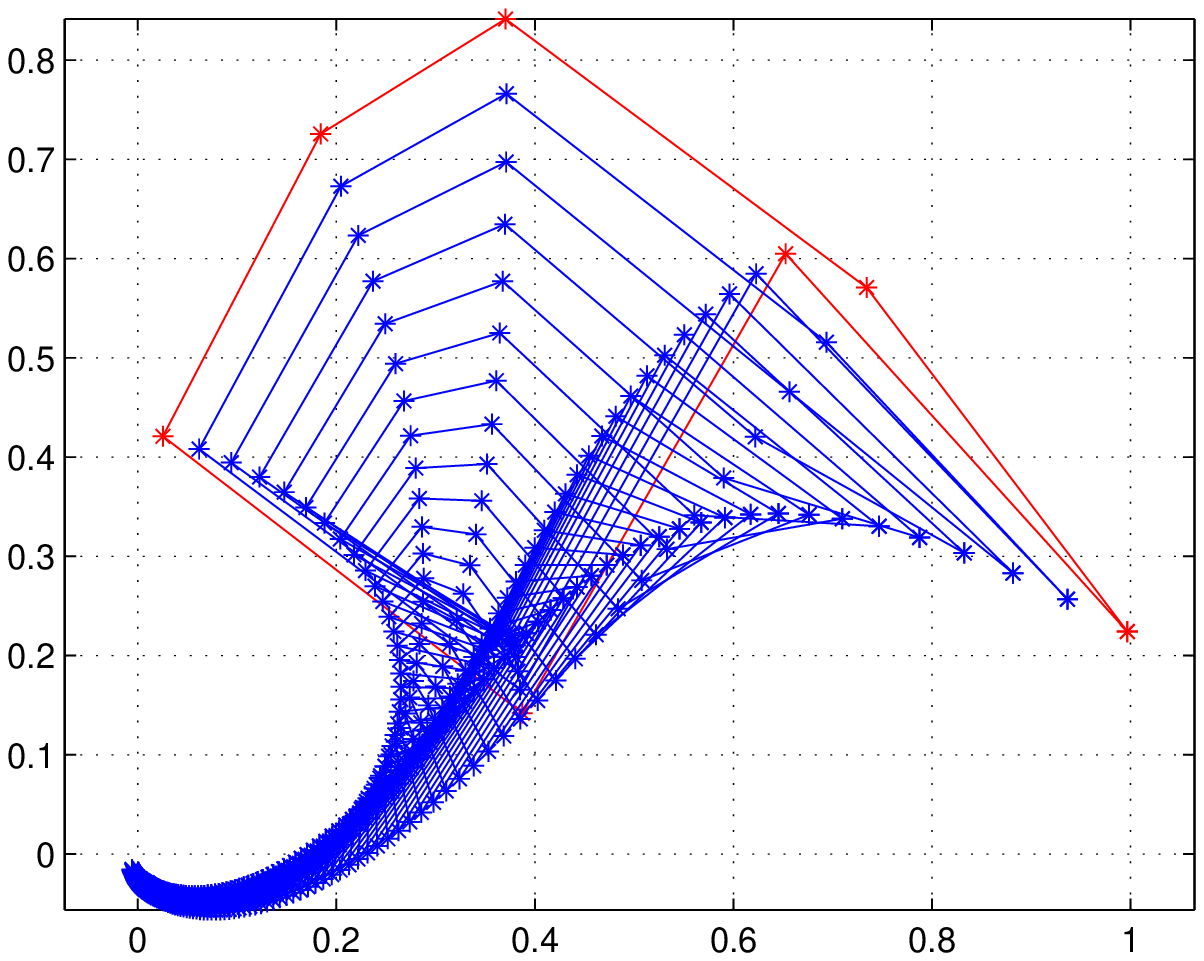}\includegraphics[scale=0.4]{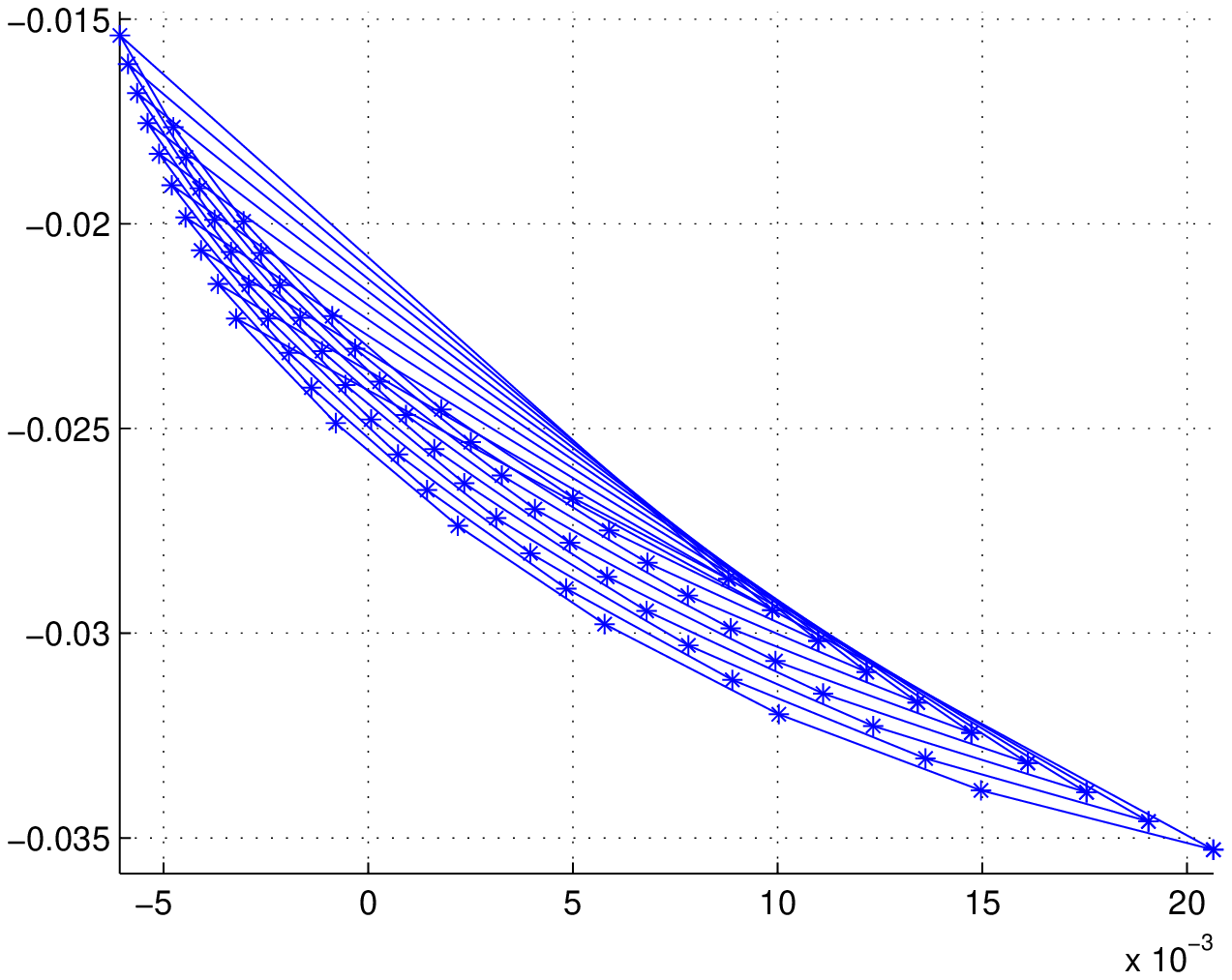}
\caption{$\lambda=\exp(i \pi/4)/2$, $\alpha=0.1$, 100 iterations}
\label{fig:lambdacomp}
\end{figure}

\begin{figure}
\hspace{-1cm}
\includegraphics[scale=0.4]{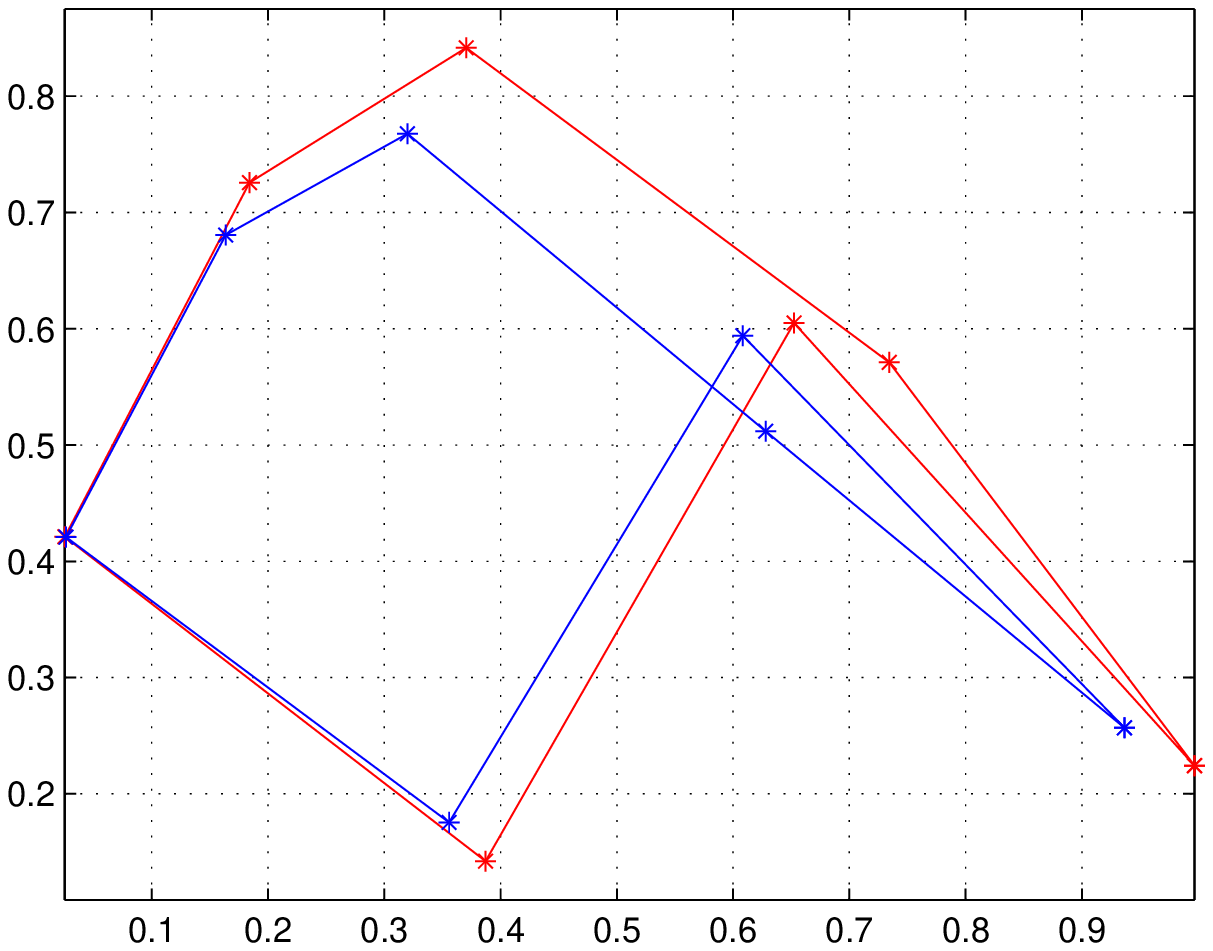}\includegraphics[scale=0.4]{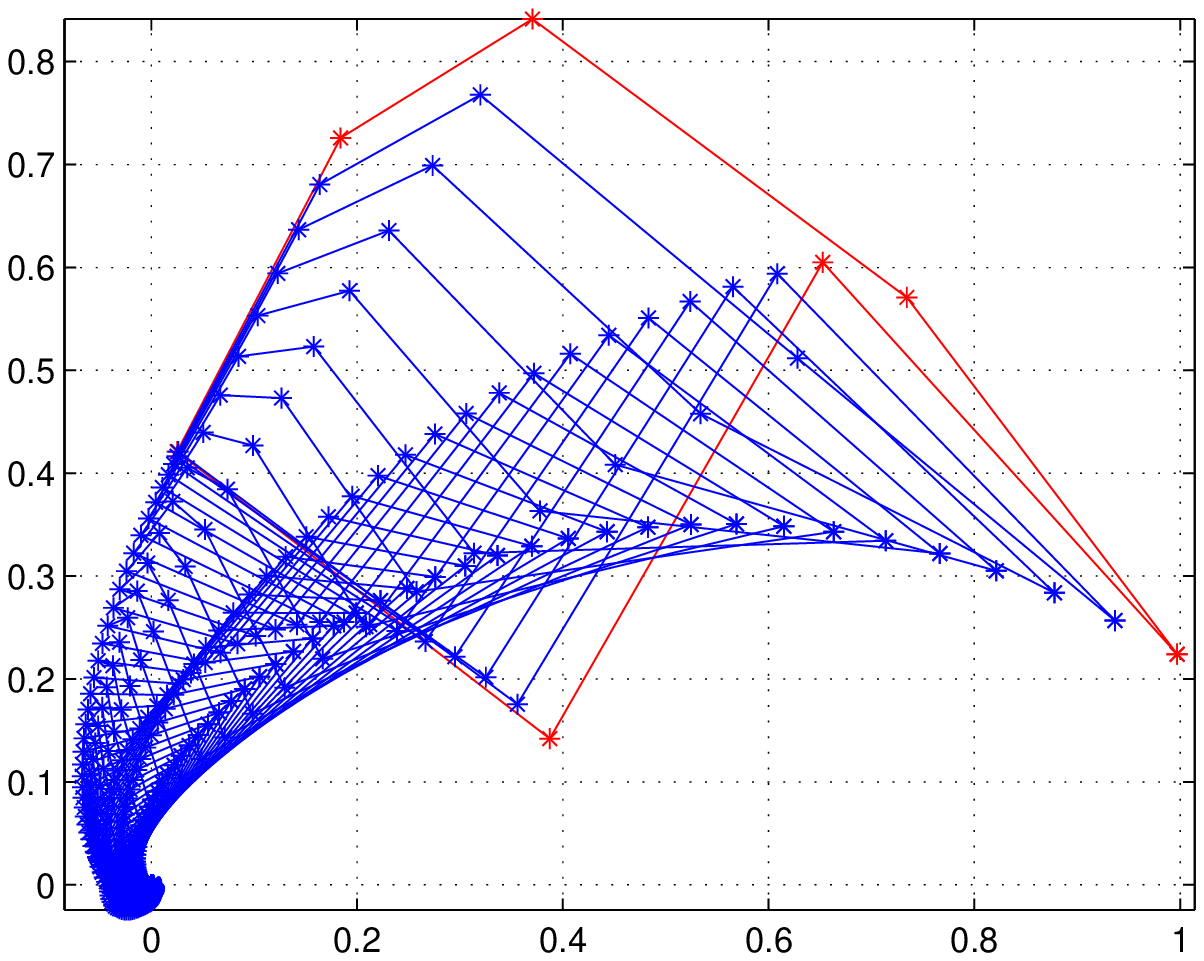}\includegraphics[scale=0.4]{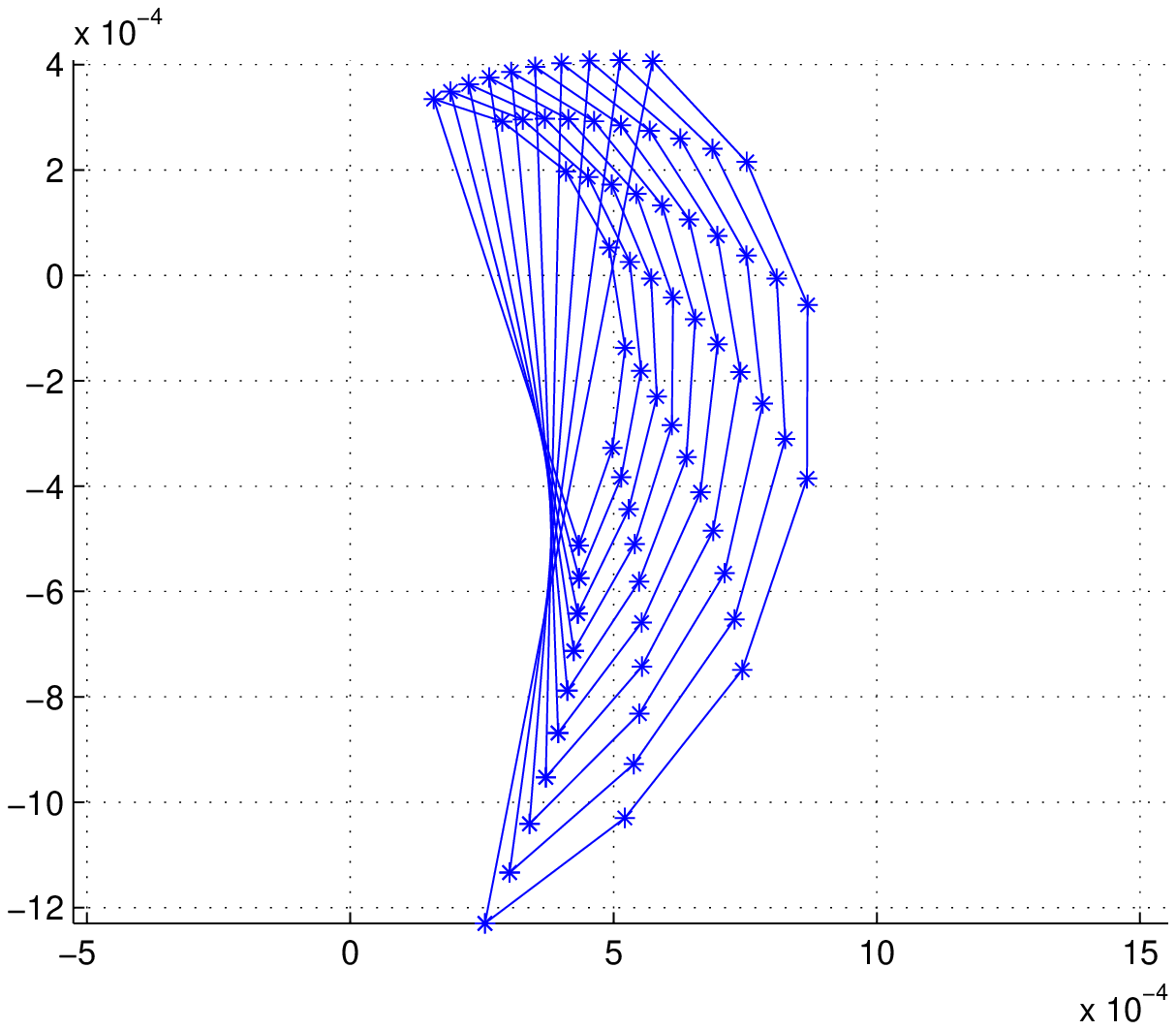}
\caption{$\lambda=\exp(i \pi/4 +i\pi)/2$, $\alpha=0.1$, 100 iterations}
\label{fig:lambda_comp}
\end{figure}

Next suppose we have $\beta_B\neq \beta_F$. Then we have 
a $\lambda=\beta_B/\beta_F$ factor circulant and the modes of the 
$\widetilde{\Pm}(t)$ evolution is controlled by
\[
\mu_l=\alpha+\sum_{k=1}^{N-1}\beta_F\left(\frac{\beta_B}{\beta_F}\right)^{k/N}
e^{-i\frac{2\pi}{N}kl},~l=0,\ldots,N-1.
\]
Here
\begin{eqnarray*}
\mu_0 
&=& 
\alpha -\beta_F 
+\beta_F\sum_{k=0}^{N-1}\left(\frac{\beta_B}{\beta_F}\right)^{k/N} \\
&=& 
\alpha-\beta_F+\beta_F\frac{\beta_B/\beta_F-1}{(\beta_B/\beta_F)^{1/N}-1}\\
&=& 
\alpha-\frac{1-\alpha}{N-1}+
\left(\frac{1-\alpha}{N-1}\right)
\left(\frac{\lambda-1}{\lambda^{1/N}-1}\right).
\end{eqnarray*}
Similarly we have that
\begin{eqnarray*}
\mu_l
&=&\alpha+\sum_{k=1}^{N-1}\beta_F
\left(\frac{\beta_B}{\beta_F}\right)^{k/n}e^{-i\frac{2\pi}{N}kl}\\
&=& \alpha-\frac{1-\alpha}{N-1}+
\left(\frac{1-\alpha}{N-1}\right)
\left(\frac{\lambda e^{-i\pi l}-1}{\lambda^{1/N}e^{i\pi l/N}-1}\right).
\end{eqnarray*}
In this example too, as before, we have
\[
\Pm(t)_{t\rightarrow\infty}
=
\left[
\begin{array}{l}
1~~~\\
~~~\lambda^{1/N}\\
~~~~~\ddots\\
0~~~~~~~~ \lambda^{\frac{N-1}{N}}
\end{array}
\right]
[\mathbf{FT}]
\left[
\begin{array}{cccc}
\mu_0^t &&&\\
 & \mu_1^t &&\\
&&\ddots &\\
0 &&& \mu_{N-1}^t
\end{array}
\right]
[\mathbf{FT}]^*
\left[
\begin{array}{l}
1\\
~~~\lambda^{-1/N} \\
~~~~~\ddots \\
0~~~~~~~ \lambda^{\frac{-(N-1)}{N}}
\end{array}
\right]
{\Pm}(0)
\]
and if $\mu_0$ is the dominant eigenvalue, we shall have 
\[
\Pm(t)_{t\rightarrow\infty}
=
\mu_0^t
\frac{1}{N}
\left[
\begin{array}{c}
1\\
\lambda^{1/N}\\
\vdots\\
\lambda^{N-1/N}
\end{array}
\right]
[
1,\lambda^{-1/N},\ldots,\lambda^{-(N-1)/N}
]
\Pm(0)
\]
and depending on the values selected for $\lambda$, we can get 
a wealth of interesting behaviors while the solutions converge or 
diverge to infinity, displaying spiralling or in line marching.
See Figures \ref{fig:lambda01}, \ref{fig:lambda_1}, \ref{fig:lambdai}, 
\ref{fig:lambda_i}, \ref{fig:lambdacomp}, \ref{fig:lambda_comp} 
where we present a few interesting cases.

%
%

\section{Concluding Remarks}

We discussed in this paper a special type of cyclic multiagent
interaction modeled by $\lambda$-factor cyclic matrices. Such
matrices allow explicit closed form diagonalizations via generalized
Fourier transforms hence enable the analysis of the evolution of
the swarm via a nice, geometric, modal decomposition process. It is
expected that a wealth of further similar, structured and nearly
cyclic interactions will also yield explicit closed form solutions
for their asymptotic behavior.
In fact, we may use evolutions that fix one, two \cite{WB} or 
several agents in the swarm and use circulant or $\lambda$-circulant 
interactions for the rest of them leading to further highly  
structured matrices that can be diagonalized, and correspondingly leading 
to interesting and explicitly predictable and designable swarm dynamics. 
In closing, we note that Turing's morphogenesis may be regarded 
as a further example of such dynamics for points in the plane where the $x$ 
and the $y$ coordinates are subjected to different linear circulant 
transformations also readily generalizable to $\lambda$-circulant maps 
\cite{T}. An analysis of such 
swarm interaction for multiagent system is forthcoming.

%
%

\section*{Acknowledgments}

The work of Fr\'ed\'erique Oggier is supported in part by the Nanyang
Technological University under Research Grant M58110049.

The work of Alfred Bruckstein is supported in part by a Nanyang Technological 
University visiting professorship, at the SPMS and IMI center.

%
%

\bibliographystyle{plain}
\bibliography{refs}
\end{document}